\documentclass[aps,prl,reprint,superscriptaddress]{revtex4-2}
\usepackage{amssymb}
\usepackage{amsbsy}
\usepackage{amsmath}
\usepackage{amsthm}
\usepackage{graphicx}
\usepackage{graphics}
\usepackage{setspace}
\usepackage{array}
\usepackage{color}
\usepackage{xcolor}
\usepackage{fontenc}
\usepackage{textcomp}
\usepackage{bm}
\usepackage{pifont}
\usepackage{soul}
\usepackage[T1]{fontenc}
\usepackage{mathdots}

\usepackage{dsfont}
\usepackage{bbold}

\newcommand{\n}{\nonumber}

\definecolor{darkred}{rgb}{0.8,0,0}
\definecolor{royalblue}{rgb}{0.0, 0.14, 0.4}
\definecolor{magenta}{cmyk}{0,.9,0,0.2}
\definecolor{amethyst}{rgb}{0.6, 0.4, 0.8}
\definecolor{cadmiumgreen}{rgb}{0.0, 0.42, 0.24}
\definecolor{deepcarmine}{rgb}{0.66, 0.13, 0.24}
\definecolor{forestgreen}{rgb}{0.13, 0.55, 0.13}

\definecolor{riceblue}{RGB}{0,32,91}
\definecolor{ricegray}{RGB}{124,126,127}
\definecolor{lightbluegray}{RGB}{173, 199, 220}
\definecolor{lightgray}{RGB}{224, 226, 230}
\definecolor{brightblue}{RGB}{159, 221, 249}
\definecolor{mediumblue}{RGB}{77, 154, 212}
\definecolor{richblue}{RGB}{10, 80, 158}
\definecolor{midnightblue}{RGB}{19, 19, 62}
\definecolor{darkgray}{RGB}{68, 71, 79}
\definecolor{darkbluegray}{RGB}{48, 59, 97}
\definecolor{warmyellow}{RGB}{233, 161, 57}
\definecolor{brickred}{RGB}{192, 72, 41}
\definecolor{burgundy}{RGB}{104, 19, 46}
\definecolor{shadowpurple}{RGB}{54, 46, 82}
\definecolor{leafemerald}{RGB}{0, 91, 80}
\definecolor{treegreen}{RGB}{0, 67, 44}
\definecolor{grassgreen}{RGB}{53, 146, 69}
\definecolor{brightgreen}{RGB}{165, 193, 81}

\usepackage[bookmarks=false,linkcolor=blue,urlcolor=blue,colorlinks,citecolor=blue]{hyperref}

\newcommand{\blue}[1]{{\color{blue}{#1}}}

\newcommand{\bsub}{\begin{subequations}}
\newcommand{\esub}{\end{subequations}}

\newcommand{\titlename}{Quantum Fisher information of magnetic quantum phase transition on Kondo lattice}

\begin{document}
\title{\titlename}
\author{Yuan Fang}
\affiliation{Department of Physics and Astronomy,  Extreme Quantum Materials Alliance, Smalley Curl Institute, Rice University, Houston, Texas 77005, USA}

\author{Lei Chen}
\affiliation{Department of Physics and Astronomy,  Extreme Quantum Materials Alliance, Smalley Curl Institute, Rice University, Houston, Texas 77005, USA}
\affiliation{Department of Physics and Astronomy, Stony Brook University, Stony Brook, NY 11794, USA}

\author{Mounica Mahankali}
\affiliation{Department of Physics and Astronomy,  Extreme Quantum Materials Alliance, Smalley Curl Institute, Rice University, Houston, Texas 77005, USA}

\author{Fang Xie}
\affiliation{Department of Physics and Astronomy,  Extreme Quantum Materials Alliance, Smalley Curl Institute, Rice University, Houston, Texas 77005, USA}

\author{Yiming Wang}
\affiliation{Department of Physics and Astronomy,  Extreme Quantum Materials Alliance, Smalley Curl Institute, Rice University, Houston, Texas 77005, USA}

\author{Shouvik Sur}
\affiliation{Department of Physics and Astronomy,  Extreme Quantum Materials Alliance, Smalley Curl Institute, Rice University, Houston, Texas 77005, USA}

\author{Qimiao Si}
\affiliation{Department of Physics and Astronomy,  Extreme Quantum Materials Alliance, Smalley Curl Institute, Rice University, Houston, Texas 77005, USA}

\begin{abstract}
Strange metals exemplify highly collective quantum many-body systems that call for new means of characterization, and there is considerable potential for quantum information approaches contributing to the cause. We investigate multipartite entanglement across the quantum phase transition of a Kondo lattice model using the quantum Fisher information (QFI). We show that the QFI associated with the spin components transverse to the order parameter characterizes the destruction of heavy quasiparticles in the Kondo-destroyed magnetic-ordered phase. The physical origin of this observation is elucidated through an analysis of the antiferromagnetic Heisenberg model. We propose to test the results in terms of both unpolarized and polarized inelastic neutron scattering measurements in the ordered part of the heavy fermion phase diagram. Our findings illustrate how different operators of a many-body system can be employed to not only witness multipartite entanglement in different sectors and but also elucidate the overall physics across different parts of the phase diagram.
\end{abstract}

\maketitle

\blue{\emph{Introduction}}---
Strange metals arise in a variety of strongly correlated systems~\cite{Kei17.1,paschen2021quantum}.
A canonical setting is provided by quantum critical heavy fermion metals.
Here the physics of Kondo destruction characterizes the loss of heavy quasiparticles~\cite{hu2022quantumc,paschen2021quantum,Coleman2001,Senthil2004}.
Studies in these systems showcase the emerging profile for strange metallicity that goes beyond the hallmark linear-in-$T$ dependence of the electrical resistivity.
It includes a transformation of the Fermi surface from large to small as a quantum critical point is crossed, a loss of quasiparticles everywhere on the Fermi surface when the system is located at the quantum critical point, and a dynamical scaling in spin and charge responses~\cite{hu2022quantumc,Si2001,Coleman2001,Senthil2004,paschen2021quantum,stefan2020}.
This rich behavior highlights quantum criticality that goes beyond the Landau paradigm, and calls for novel means of characterization. 
Recent studies have shown that entanglement witnesses, particularly quantum Fisher information (QFI), provide a promising way for this cause~\cite{Fang2025Amplified,Mazza2026Quantum}.

Entanglement refers to correlations in quantum systems that go beyond any classical description~\cite{bell2004speakable,zeng2019quantum}.
It is a central resource for quantum information processing and quantum computation~\cite{nielsen2001quantum,Chitambar2019Quantum,Tan2021Fisher}.
A widely used way to quantify entanglement is via bipartite measures such as the entanglement entropy, which has been extensively studied in a broad range of quantum systems~\cite{amico2008entanglement,kitaev2006topological,li2008entanglement,Wolf2008Area,Toth2012Multipartite}.
However, bipartite measures are limited in their ability to characterize multipartite entanglement and are often challenging to access experimentally.
Recently, QFI has been proposed as an experimentally accessible witness of multipartite entanglement in quantum many-body systems~\cite{hyllus2012fisher,Pezze2017Multipartite,Pezze2018Quantum,hauke2016measuring}. 
QFI depends on both the quantum state and a chosen set of local Hermitian operators, which define a multi-partition of the Hilbert space. 
As a result, QFI detects entanglement among these partitions and provides a lower bound on the entanglement depth, defined as the minimal number of subsystems that are genuinely entangled~\cite{hyllus2012fisher}. 
An important advantage of QFI is that it can be directly extracted from spectroscopy of the corresponding operators, making it experimentally measurable~\cite{hauke2016measuring}. 
In addition to the heavy fermion strange metals~\cite{Fang2025Amplified,Mazza2026Quantum}, QFI is of interest to diagnosing entanglement in a variety of condensed matter systems~\cite{George2020Experimental,Pratt2022Spin,Zhang2022Multipartite,Scheie2021Witnessing,*Scheie2023Erratum,scheie2023proximate,Laurell2021Quantifying,hales2023witnessing,Lambert2020Revealing,Liu2024Entanglement,Baykusheva2023Witnessing,Balut2024Quantum,BALUT20251354750,Ji2025Density,Guan2025Exploring,Scheie2024Tutorial,Liu2020Quantum,Adesso_2016}.

\begin{figure}[t]
    \centering
    \includegraphics[width=\linewidth]{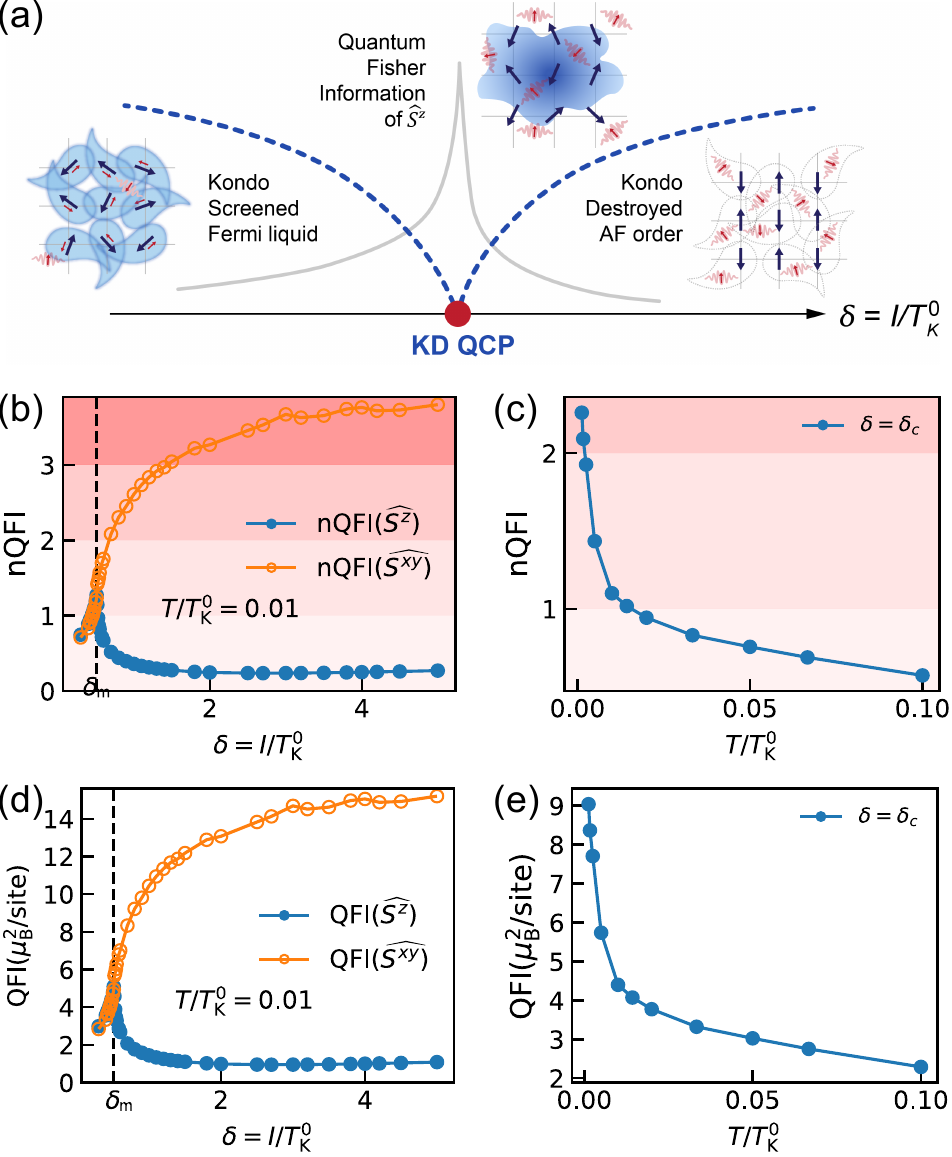}
    \caption{
    (a) Quantum Fisher information of the operator $\widehat{S^z}$ reveals amplified entanglement at the Kondo destruction (KD) QCP, which separates a Kondo-screened Fermi-liquid phase and a Kondo-destroyed AF-ordered phase. Here the spin direction $z$ is the direction of the magnetic order parameter. The control parameter $\delta=I/T_{\mathrm K}^0$ is the ratio between the RKKY interaction and Kondo temperature. Adapted from Ref.\,\cite{Fang2025Amplified}.
    (b) Normalized QFI density (nQFI) of the longitudinal $\widehat{S^z_i}$ and transverse $\widehat{S^x_i}$(or $\widehat{S^y_i}$) spin operators as a function of the control parameter $\delta=I/T_{\mathrm{K}}^0$, where $I$ is the RKKY interaction and $T_{\mathrm{K}}^0$ is Kondo temperature. Here the temperature is fixed at a relatively low value, $T/T_{\mathrm{K}}^0 = 0.01$. The dashed vertical line indicates the magnetic phase transition at $\delta_m=I_m/T_{\mathrm{K}}^0 \approx 0.51$.
    (c) nQFI at the quantum critical coupling $\delta_c=I_c/T_{\mathrm{K}}^0 \approx 0.47$.
    (d),(e) The un-normalized QFI density of of the AF magnetization operator, $f_Q(g\mu_{\mathrm{B}})^2$, suitable for comparison with inelastic neutron scattering experiments. Here $g=2$ is used.
    }
    \label{fig:Fig_scheme}
\end{figure}

In the Kondo lattice, QFI can characterize the Kondo-destruction quantum critical point (KD QCP)~\cite{Fang2025Amplified}; the latter separates the Kondo-screened Fermi-liquid phase and the Kondo-destroyed antiferromagnetic (AF) phase~\cite{Si2001,Coleman2001,Senthil2004,paschen2021quantum,stefan2020,Coleman_review,hu2022quantumc,Liyang-Chen2023}.
Specifically, the QFI constructed from local moment spin operators $\widehat{S^z}$ was found to peak at KD QCP, showing enhanced multipartite entanglement associated with Kondo destruction.
The enhanced QFI at QCP implies the breakdown of Kondo entanglement and therefore characterizes the loss of heavy quasiparticles in strange metals.
This QFI decreases drastically when the system enters the Kondo-screened phase or the Kondo-destroyed phase.
A schematic phase diagram drawn together with the behavior of this QFI is shown in Fig.~\ref{fig:Fig_scheme}(a).

A natural question then arises: 
if Kondo destruction at the QCP is seen by the spin QFI, is the Kondo destruction on the other side of the QCP also seen by a spin QFI?
This question is nontrivial since previous work showed that QFI associated with any local spin operators --in contrast to some nonlocal operators -- becomes trivial in the Fermi-liquid phase~\cite{wang2025local}.
However, the Kondo-destroyed AF-ordered phase of the Kondo lattice represents a distinct ground state that exhibits long-range order and gapless collective modes.
Whether the QFI can reveal multipartite entanglement that, in turn, characterize the destruction of heavy quasiparticles, in this phase remains an open question.

In this work, we show that the QFI of the transverse spin operators captures multipartite entanglement in the AF-ordered phase.
By studying the Kondo lattice model with extended dynamical mean-field theory (EDMFT)~\cite{hu2022quantumc,Si1996,Smith2000,Chitra2000,hu2022extended}, we demonstrate that the QFI associated with the longitudinal spin operator $\widehat{S^z}$ becomes trivial (i.e., not witnessing any multipartite entanglement) in the AF-ordered phase, while the QFI associated with the transverse spin operator $\widehat{S^x}$ or $\widehat{S^y}$ is nontrivial, revealing multipartite entanglement in the ordered state.
To further elucidate this mechanism, we use linearized spin-wave theory to analyze the spin QFIs in a quasi-2D Heisenberg model~\cite{manousakis1991rmp,auerbach2012interacting,Sandvik2010}. 
This analysis shows behavior of the spin QFI similar to that seen in the AF-ordered phase of the Kondo lattice obtained from the EDMFT results. 
We attribute the anisotropic behavior to the interplay between quantum and thermal fluctuations in the broken symmetry state.
Our proposal of the transverse spin QFI witnessing multipartite entanglement in the ordered phase of the Kondo lattice model is experimentally testable using a combination of unpolarized and polarized neutron scattering measurements and promises a quantum information means to elucidate the physics of strongly correlated electronic states in magnetic settings.

\blue{\emph{Kondo lattice}}---
The Kondo lattice model provides a prototypical platform for studying the competition between distinct forms of quantum entanglement in strongly correlated electronic systems.
In this model, local magnetic moments are coupled both to itinerant conduction electrons and to the other local moments, through Kondo coupling and Ruderman--Kittel--Kasuya--Yosida (RKKY) interaction respectively, giving rise to two competing entanglement mechanisms.
On the one hand, Kondo coupling favors the formation of local Kondo singlets, entangling the local moments with the conduction electron;
on the other hand, RKKY interaction promotes entanglement among the local moments, leading to AF-ordered states.
The competition between these two types of entanglement generates a rich phase diagram~\cite{Si2001,Coleman2001,Senthil2004,hu2022quantumc}.
Tuning the RKKY interactions can drive the system through the KD QCP, in the vicinity of  which extreme quantum fluctuations destroys the Kondo screening and give rise to strange metallicity.
This regime is characterized by anomalous dynamical scaling and absence of well-defined quasiparticles, making the Kondo lattice model an ideal setting for investigating how multipartite entanglement evolves across quantum phase transitions in strongly correlated systems~\cite{Si2001,hu2022quantumc}.

To describe this physics, we consider the $\mathrm{SU(2)}$ periodic Anderson lattice model with the Hamiltonian
\begin{equation}
\begin{aligned}
    H &= \frac{U}{2}\sum_i \Big[\sum_{\sigma} f_{i\sigma}^\dagger f_{i\sigma} -1\Big]^2 + \sum_{i,\sigma} \epsilon_f f_{i\sigma}^\dagger f_{i\sigma} \\
    &\quad + V\sum_{i,\sigma} \left(c_{i\sigma}^\dagger f_{i\sigma}
    + f_{i\sigma}^\dagger c_{i\sigma}\right)
    + \sum_{p,\sigma} \epsilon_k\, c_{k\sigma}^\dagger c_{k\sigma} \\ 
    &\quad+ \sum_{ij} I_{ij}\, \mathbf{S}_i \cdot \mathbf{S}_j \, .
\end{aligned}
\label{eq:AM}
\end{equation}
Here, $f_{i\sigma}^\dagger$ ($c_{i\sigma}^\dagger$) creates a localized ($f$) electron or a conduction ($c$) electron with spin $\sigma$ on site $i$.
The on-site Coulomb repulsion $U$ enforces filling constraints on the $f$ electrons and generates effective local moments at low-energy sectors.
The hybridization $V$ couples the local moments to the itinerant conduction electrons with dispersion $\epsilon_k$ and gives rise to Kondo screening. 
The local moment spin operator is defined as $\mathbf{S}_i = f_i^\dagger \boldsymbol{\sigma} f_i /2$.
The term $I_{ij}$ describes the AF RKKY interaction between local moments, which favors AF ordering.
We consider an RKKY interaction whose Fourier transform $I_{\mathbf{q}} = I(\cos q_x + \cos q_y)$ is minimized at the AF ordering vector $\mathbf{Q}=(\pi,\pi)$.

We solve the model by EDMFT, in which correlation functions of the lattice model are calculated from a self-consistent Bose-Fermi Anderson (BFA) impurity model.
In this formulation, the local $f$ electron couples to a fermionic bath and a bosonic bath that encode the effects of Kondo hybridization and RKKY interactions, respectively~\cite{hu2022quantumc,hu2022extended}.

\blue{\emph{Quantum Fisher information of spin operators}}---
QFI is an entanglement witness.
For an $m$-producible state and an operator $\widehat{O}=\sum_{i=1}^N \widehat{O_i}$ constructed from local operators $\widehat{O_i}$ with spectrum in the interval $[h_{\mathrm{min}},h_{\mathrm{max}}]$, the QFI density $f_Q=F_Q/N$ is bounded above by $f_Q \leq m\,(h_{\mathrm{max}}-h_{\mathrm{min}})^2$~\cite{hyllus2012fisher,Toth2012Multipartite}.
The dimensionless entanglement witness, normalized QFI density, is defined as $\mathrm{nQFI}=f_Q/(h_{\mathrm{max}}-h_{\mathrm{min}})^2$.

We determine the nQFIs for the periodic Anderson lattice model and spin operators $\widehat{O^a} = \sum_{j=1}^N  e^{i\mathbf{Q}\cdot \mathbf{r}_i} \widehat{S^a_i}$, where $a=x,y,z$ and $\mathbf{Q}$ is the AF ordering vector.
Fig.~\ref{fig:Fig_scheme}(b) shows nQFI of the spin operators $\widehat{S^z_i}$ and $\widehat{S^x_i}$(or $\widehat{S^y_i}$) as a function of the control parameter $\delta=I/T_{\mathrm{K}}^0$.
In the paramagnetic regime ($\delta < \delta_m$), the system preserves $\mathrm{SU(2)}$ spin rotation symmetry.
Consequently, the QFI densities for different spin components are identical.
In this Kondo-screened heavy-Fermi-liquid phase, the nQFI remains trivial (nQFI$\leq 1$) and does not provide a multipartite entanglement witness for these local operators, consistent with previous analyses~\cite{wang2025local}.
The nQFI at KD QCP is shown in Fig.~\ref{fig:Fig_scheme}(c), where the nQFI does not saturate even down to $T/T_{\mathrm{K}}^0=0.00125$ (note $SU(2)$ symmetry is also preserved at QCP).
In Figs.~\ref{fig:Fig_scheme}(d,e) we show the un-normalized QFI density corresponding to Figs.~\ref{fig:Fig_scheme}(b,c), for comparison with potential future inelastic neutron scattering experiments.

In the Kondo-destroyed AF phase ($\delta > \delta_m$), the spin rotation symmetry is spontaneously broken.
In this phase, the longitudinal $\mathrm{nQFI}(\widehat{S^z})$ remains trivial, while the transverse $\mathrm{nQFI}(\widehat{S^{xy}})$ gets enhanced. 
The reduction in longitudinal nQFI originates from the ordering of local moments and the suppression of longitudinal fluctuations. 
The enhancement in transverse nQFI originates from the emergence of gapless Goldstone modes, which give rise to large transverse quantum fluctuations.
Such behavior is consistent with general expectations for magnetically ordered systems with spontaneously broken continuous symmetries, as we show in the next section. 
We note that Kondo correlations can persist inside the AF-ordered phase and evolve continuously across the QCP.
In the ordered phase, transverse nQFI increases when $I$ increases. This is consistent with the coexistence of strong transverse spin fluctuations with gradually weakening Kondo correlations.

\begin{figure}[t]
    \centering
    \includegraphics[width=\linewidth]{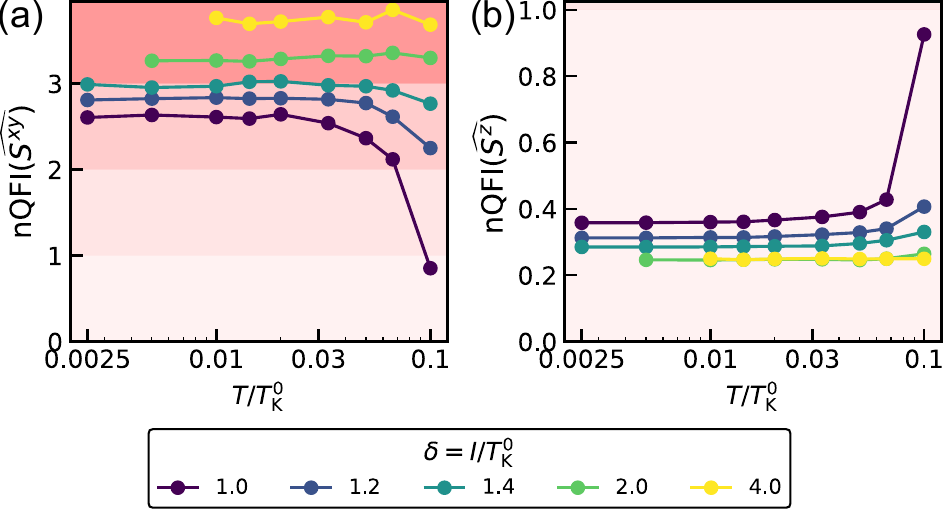}
    \caption{
    (a)~Normalized QFI density (nQFI) of transverse spin operator $\widehat{S^x_i}$(or $\widehat{S^y_i}$) as a function of temperature at different values of the control parameter $\delta=I/T_{\mathrm{K}}^0$ in the AF-ordered part of the phase diagram [cf. Fig.~\ref{fig:Fig_scheme}(a)]; 
    (b)~nQFI of longitudinal spin operator $\widehat{S^z_i}$ as a function of temperature at different $\delta$.
    }
    \label{fig:edmft_T}
\end{figure}

Figs.~\ref{fig:edmft_T}(a) and (b) show the temperature dependence of the transverse and longitudinal nQFI for several values of $\delta$.
These values of $\delta$ are chosen such that the system stays within the AF-ordered phase even at the highest temperature considered.
The spin nQFIs saturate at low temperatures, thereby displaying some characteristic temperature scale in the entanglement depth.
This is in contrast to their behavior at the KD QCP, where no saturation temperature scale is observed [see Fig.~\ref{fig:Fig_scheme}(c)].
As temperature increases, thermal fluctuations progressively suppress the coherent transverse correlations; accordingly, the transverse QFI monotonically decreases with temperature.
Thermal fluctuations also suppress the magnetic order and release fluctuations in the longitudinal channel, leading to a slight increase of the longitudinal QFI.

\begin{figure}[t]
    \centering
    \includegraphics[width=\linewidth]{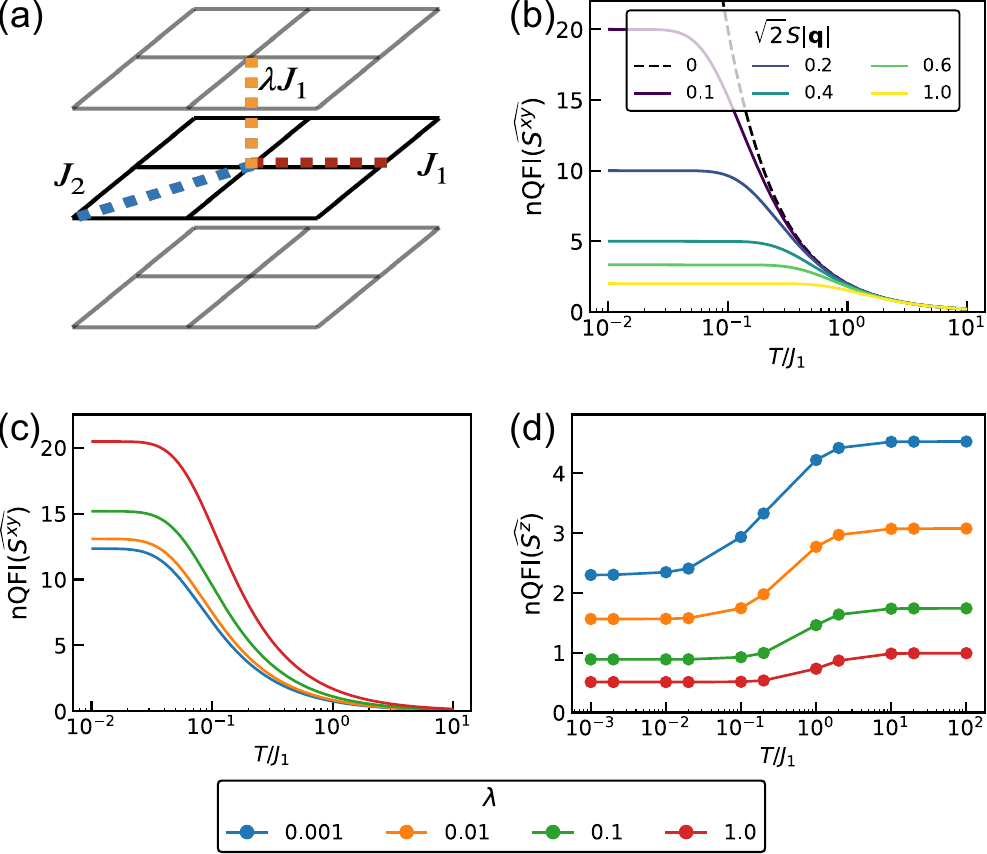}
    \caption{(a) Schematic of the quasi-2D Heisenberg model.
    (b) nQFI of $\widehat{S^x_i}$ as a function of temperature with different $\sqrt{2}S|{\mathbf{q}}|$. We set parameter $\lambda=0$ and $J_2=0$. 
    nQFI of (c) $\widehat{S^x_i}$ and (d) $\widehat{S^z_i}$ as functions of temperature for the quasi-2D Heisenberg model with different parameter $\lambda$. We set $q_\parallel=q_z$, $\sqrt{2}Sq_\parallel=0.1$ and $J_2=0$. 
    }
    \label{fig:sw_T}
\end{figure}

\blue{\emph{Origin of enhanced transverse quantum Fisher information}}---
To gain further insight into the origin of the enhanced transverse spin QFI in the AF-ordered phase, we analyze a quantum Heisenberg model using linearized spin wave theory.
Specifically, we consider the quasi-2D tetragonal lattice $J_1$-$J_2$ Heisenberg model with AF couplings. It is described by the Hamiltonian:
\begin{equation}
    H = \sum_{\langle i,j \rangle} J_{ij} \mathbf{S}_i \cdot \mathbf{S}_j + J_2 \sum_{\langle\langle i,j \rangle\rangle} \mathbf{S}_i \cdot \mathbf{S}_j \,,
\end{equation}
where $\langle i,j \rangle$ stands for both in-plane and out-of-plane nearest terms and $\langle\langle i,j \rangle\rangle$ only stands for the in-plane next-nearest neighbors.
We set $J_{ij}=J_1$ for in-plane nearest neighbors and $J_{ij}=\lambda J_1$ for out-of-plane nearest neighbors.
For sufficiently small anisotropic parameter $\lambda$, the Hamiltonian describes a 2D Heisenberg model. 
A schematic of this model is shown in Fig.~\ref{fig:sw_T}(a).
For $\lambda\neq 0$, long-range AF order may appear at nonzero temperature~\cite{MerminWagner,manousakis1991rmp}.
This model exhibits a N\'{e}el AF-ordered phase with ordering wave vector $\mathbf{Q}=(\pi,\pi,\pi)$ for $J_2/J_1 \lesssim 0.5$, while stronger frustration leads to nonmagnetic or nematic phases at larger $J_2/J_1$.
Here we focus on the N\'{e}el AF-ordered phase.

Using the spin wave theory, we find the nQFI associated with the transverse spin operators is given by~\cite{sm}:
\begin{align}\label{eqn:fQ_sw_xy}
    \mathrm{nQFI}(\widehat{S^{xy}},\mathbf{q},T) = \frac{z J_1}{\omega_{\mathbf q}} 
    \left(\frac{\bar S}{S}\right)^2 \tanh\left( \frac{\beta \omega_{\mathbf q}}{2} \right) \,,
\end{align}
where $z=4+2\lambda$ is the effective coordinate number, $\beta=1/k_{\mathrm{B}}T$, $\omega_{\mathbf q}$ is the dispersion of the gapless spin wave, $\mathbf q$ is the wave vector away from ordering vector $\mathbf Q$, $S$ and $\bar S$ denote the bare and effective spin, respectively (see End Matter).

Fig.~\ref{fig:sw_T}(b) shows its temperature dependence in this 2D Heisenberg model at $\lambda=0$, $J_2=0$ for a few different $\mathbf{q}$.
For a finite system, we estimate $\mathrm{nQFI}(\widehat{S^x},0,T=0)\propto 1/|\mathbf{q}|\sim \sqrt{N}$ where $N$ is the number of sites. 
This shows that the transverse nQFI is divergent in the thermodynamic limit~\footnote{Conventionally, spontaneous symmetry breaking occurs only in the thermodynamic limit for a lattice model. Here, however, we solve the lattice model around a symmetry-broken ordered state, so the thermodynamic limit can be taken afterward.}.

In the unfrustrated limit $J_2=0$, the model reduces to the standard AF Heisenberg model. 
In this regime, the transverse nQFI behavior provides a useful point of comparison for the AF-ordered phase of the Kondo lattice at large RKKY interaction strength.
Fig.~\ref{fig:sw_T}(c) shows the temperature dependence of transverse nQFI for our model with different anisotropy parameter $\lambda$, which is robust for finite $\lambda$.
The same qualitative trend is seen in the EDMFT results in Fig.~\ref{fig:edmft_T}(a) where transverse nQFI saturates upon cooling and decreases monotonically as temperature increases.

At zero temperature, Eq.~(\ref{eqn:fQ_sw_xy}) 
diverges as $\mathbf{q} \to 0$ due to the presence of gapless Goldstone modes.
Moreover, as $J_2/J_1$ approaches $1/2$, the spin-wave velocity vanishes and the transverse nQFI is strongly enhanced, reflecting the softening of the Goldstone mode near the instability of N\'{e}el order.

The longitudinal nQFI is contributed by two-magnon processes~\cite{sm}.
For $\mathbf q= 0$ it still remains finite:
\begin{align}\label{eqn:fQ_sw_z}
    \mathrm{nQFI}(\widehat{S^z},0,T) &= \frac{1}{S^2} \int_{BZ} \frac{d^3k}{(2\pi)^3} \frac{1}{1/\gamma_{\mathbf k}^2 - 1} \frac{\tanh(\beta \omega_{\mathbf k})}{\tanh(\beta \omega_{\mathbf k}/2)} \,, 
\end{align}
where the integral is over the Brillouin zone and $\gamma_{\mathbf k}=\frac{1}{z} \left[2\cos k_x + 2\cos k_y + 2 \lambda \cos k_z\right]$.
The longitudinal spin QFI density at zero temperature and that at infinite temperature have a simple relation, $2\times \mathrm{nQFI}(\widehat{S^z},0,T=0) = \mathrm{nQFI}(\widehat{S^z},0,T=\infty)$.
Fig.~\ref{fig:sw_T}(d) shows the temperature dependence of longitudinal nQFI for the quasi-2D Heisenberg model with different anisotropy parameter $\lambda$.
The same qualitative trend is seen in the EDMFT results in Fig.~\ref{fig:edmft_T}(a) where longitudinal nQFI saturates upon cooling and increases monotonically as temperature increases.

\blue{\emph{Discussion and Conclusion}}---
In this work, we investigate multipartite entanglement across the magnetic quantum phase transition of the Kondo lattice model.
We find that the QFI constructed from transverse spin operators remains nontrivial throughout the AF-ordered phase, revealing robust multipartite entanglement even in the presence of a long range magnetic order.
This in turn impedes the entanglement pathway for the formation of Kondo singlets and, accordingly, for the development of heavy qausiparticles [cf., Fig.~\ref{fig:Fig_scheme}(a)].
The transverse spin QFI can be determined using unpolarized inelastic neutron scattering experiments, while the contrast of the multipartite entanglement between the transverse and longitudinal spin operators can be detected by polarized inelastic neutron scattering measurements.
Thus we expect that our theoretical work will motivate further experimental probes of entanglement witnesses in heavy fermion metals and related strongly correlated electron systems.

The anisotropy in spin QFI originates from spontaneous symmetry breaking and the emergence of gapless transverse Goldstone modes.
Our spin wave analysis of the quasi-2D AF Heisenberg model shows that the enhancement of transverse QFI is a generic consequence of collective low energy excitations in AF-ordered phase.
The temperature dependence in these spin QFI reflects an intriguing interplay between quantum and thermal fluctuations. 

In addition to elucidating the physics of Kondo destruction, the strong operator dependence in the AF-ordered phase suggests a general principle.
Namely, entanglement in different channels can be detected by the QFI of the corresponding operators~\cite{frerot2024symmetry}. 
The optimal operators that maximize the QFI may therefore reveal the underlying form of entanglement. 
Our work, thus, provides further impetus for developing systematic approaches to identify such optimal operators in a variety of quantum matter settings.
At the same time, elevated multipartite entanglement can be retrieved by such means as coupling to cavity photons~\cite{sur2026amplified}. Accordingly, our work reveals a new metallic setting for novel quantum functionality.

\blue{\emph{Note added}}---
At the advanced stage of preparing this manuscript, we learnt that W. Simeth, P. Laurell and A. Scheie (unpublished) have also studied the QFI in the ordered phase of a two-dimensional spin model; where there is overlap, the results are consistent.

\blue{\emph{Acknowledgement}}---
We would like to thank Meigan Aronson, Matteo Mitrano, Haoyu Hu, Kevin Ingersent, Silke Paschen, Anders Sandvik, Allen Scheie, Vaibhav Sharma, Ming Yi and Jianxin Zhu for useful discussions.
This work has been supported in part by the NSF Grant No.\ DMR-2220603, the Robert A. Welch Foundation Grant No.\ C-1411, and the Vannevar Bush Faculty Fellowship ONR-VB N00014-23-1-2870.
The majority of the computational calculations have been performed on the Shared University Grid at Rice funded by NSF under Grant No.~EIA-0216467, a partnership between Rice University, Sun Microsystems, and Sigma Solutions, Inc., the Big-Data Private-Cloud Research Cyberinfrastructure MRI-award funded by NSF under Grant No. CNS-1338099, and the Advanced Cyberinfrastructure Coordination Ecosystem: Services \& Support (ACCESS) by NSF under Grant No. DMR170109. Y.F., L.C., F.X. and Q.S. acknowledge the hospitality of the Aspen Center for Physics, which is supported by NSF grant No. PHY-2210452.

\bibliographystyle{apsrev4-2}
\bibliography{reference.bib}

@article{sm,
  journal = 	 {Supplemental material}
}

@article{sur2026amplified,
  title={Amplified response of cavity-coupled quantum-critical systems},
  author={Sur, Shouvik and Wang, Yiming and Mahankali, Mounica and Paschen, Silke and Si, Qimiao},
  journal={Nature Communications},
  volume={17},
  number={1},
  pages={4404},
  year={2026},
  publisher={Nature Publishing Group UK London},
  doi = {10.1038/s41467-026-73112-1},
  url = {https://doi.org/10.1038/s41467-026-73112-1}
}

@article{Sandvik2010,
	author = {Sandvik, Anders W.},
	doi = {10.1063/1.3518900},
	issn = {0094-243X},
	journal = {AIP Conference Proceedings},
	month = {11},
	number = {1},
	pages = {135-338},
	title = {Computational Studies of Quantum Spin Systems},
	url = {https://doi.org/10.1063/1.3518900},
	volume = {1297},
	year = {2010},
}

@article{Kei17.1,
	author = {Keimer, B. and Moore, J. E.},
	date = {2017/11/01},
	doi = {10.1038/nphys4302},
	id = {Keimer2017},
	isbn = {1745-2481},
	journal = {Nature Physics},
	number = {11},
	pages = {1045--1055},
	title = {The physics of quantum materials},
	url = {https://doi.org/10.1038/nphys4302},
	volume = {13},
	year = {2017}}

@article{Adesso_2016,
doi = {10.1088/1751-8113/49/47/473001},
url = {https://doi.org/10.1088/1751-8113/49/47/473001},
year = {2016},
month = {nov},
publisher = {IOP Publishing},
volume = {49},
number = {47},
pages = {473001},
author = {Adesso, Gerardo and Bromley, Thomas R and Cianciaruso, Marco},
title = {Measures and applications of quantum correlations},
journal = {Journal of Physics A: Mathematical and Theoretical}
}

@article{Liu2020Quantum,
doi = {10.1088/1751-8121/ab5d4d},
url = {https://dx.doi.org/10.1088/1751-8121/ab5d4d},
year = {2019},
month = {dec},
publisher = {IOP Publishing},
volume = {53},
number = {2},
pages = {023001},
author = {Jing Liu and Haidong Yuan and Xiao-Ming Lu and Xiaoguang Wang},
title = {Quantum Fisher information matrix and multiparameter estimation},
journal = {Journal of Physics A: Mathematical and Theoretical}
}

@article{Zhang2022Multipartite,
  title = {Multipartite entanglement of the topologically ordered state in a perturbed toric code},
  author = {Zhang, Yu-Ran and Zeng, Yu and Liu, Tao and Fan, Heng and You, J. Q. and Nori, Franco},
  journal = {Physical Review Research},
  volume = {4},
  issue = {2},
  pages = {023144},
  numpages = {11},
  year = {2022},
  month = {May},
  publisher = {American Physical Society},
  doi = {10.1103/PhysRevResearch.4.023144},
  url = {https://link.aps.org/doi/10.1103/PhysRevResearch.4.023144}
}

@article{Pezze2018Quantum,
  title = {Quantum metrology with nonclassical states of atomic ensembles},
  author = {Pezz\`e, Luca and Smerzi, Augusto and Oberthaler, Markus K. and Schmied, Roman and Treutlein, Philipp},
  journal = {Reviews of Modern Physics},
  volume = {90},
  issue = {3},
  pages = {035005},
  numpages = {70},
  year = {2018},
  month = {Sep},
  publisher = {American Physical Society},
  doi = {10.1103/RevModPhys.90.035005},
  url = {https://link.aps.org/doi/10.1103/RevModPhys.90.035005}
}

@article{Wolf2008Area,
  title = {Area Laws in Quantum Systems: Mutual Information and Correlations},
  author = {Wolf, Michael M. and Verstraete, Frank and Hastings, Matthew B. and Cirac, J. Ignacio},
  journal = {Physical Review Letters},
  volume = {100},
  issue = {7},
  pages = {070502},
  numpages = {4},
  year = {2008},
  month = {Feb},
  publisher = {American Physical Society},
  doi = {10.1103/PhysRevLett.100.070502},
  url = {https://link.aps.org/doi/10.1103/PhysRevLett.100.070502}
}

@article{hyllus2012fisher,
  title = {Fisher information and multiparticle entanglement},
  author = {Hyllus, Philipp and Laskowski, Wies\l{}aw and Krischek, Roland and Schwemmer, Christian and Wieczorek, Witlef and Weinfurter, Harald and Pezz\'e, Luca and Smerzi, Augusto},
  journal = {Phys. Rev. A},
  volume = {85},
  issue = {2},
  pages = {022321},
  numpages = {10},
  year = {2012},
  month = {Feb},
  publisher = {American Physical Society},
  doi = {10.1103/PhysRevA.85.022321},
  url = {https://link.aps.org/doi/10.1103/PhysRevA.85.022321}
}

@ARTICLE{Fang2025Amplified,
	author = {Fang, Yuan and Mahankali, Mounica and Wang, Yiming and Chen, Lei and Hu, Haoyu and Paschen, Silke and Si, Qimiao},
	date = {2025/03/14},
	doi = {10.1038/s41467-025-57778-7},
	id = {Fang2025},
	isbn = {2041-1723},
	journal = {Nature Communications},
	number = {1},
	pages = {2498},
	title = {Amplified multipartite entanglement witnessed in a quantum critical metal},
	url = {https://doi.org/10.1038/s41467-025-57778-7},
	volume = {16},
	year = {2025}}

@article{hauke2016measuring,
  title={Measuring multipartite entanglement through dynamic susceptibilities},
  author={Hauke, Philipp and Heyl, Markus and Tagliacozzo, Luca and Zoller, Peter},
  journal={Nature Physics},
  volume={12},
  number={8},
  pages={778--782},
  year={2016},
  publisher={Nature Publishing Group UK London},
  doi = {https://doi.org/10.1038/nphys3700},
  url = {https://www.nature.com/articles/nphys3700}
}

@article{Pezze2017Multipartite,
  title = {Multipartite Entanglement in Topological Quantum Phases},
  author = {Pezz\`e, Luca and Gabbrielli, Marco and Lepori, Luca and Smerzi, Augusto},
  journal = {Physical Review Letters},
  volume = {119},
  issue = {25},
  pages = {250401},
  numpages = {6},
  year = {2017},
  month = {Dec},
  publisher = {American Physical Society},
  doi = {10.1103/PhysRevLett.119.250401},
  url = {https://link.aps.org/doi/10.1103/PhysRevLett.119.250401}
}

@article{Lambert2020Revealing,
  title = {Revealing divergent length scales using quantum Fisher information in the Kitaev honeycomb model},
  author = {Lambert, James and S\o{}rensen, Erik S.},
  journal = {Physical Review B},
  volume = {102},
  issue = {22},
  pages = {224401},
  numpages = {11},
  year = {2020},
  month = {Dec},
  publisher = {American Physical Society},
  doi = {10.1103/PhysRevB.102.224401},
  url = {https://link.aps.org/doi/10.1103/PhysRevB.102.224401}
}

@ARTICLE{Liu2024Entanglement,
  title = {Entanglement Witness for Indistinguishable Electrons Using Solid-State Spectroscopy},
  author = {Liu, Tongtong and Xu, Luogen and Liu, Jiarui and Wang, Yao},
  journal = {Physical Review X},
  volume = {15},
  issue = {1},
  pages = {011056},
  numpages = {31},
  year = {2025},
  month = {Mar},
  publisher = {American Physical Society},
  doi = {10.1103/PhysRevX.15.011056},
  url = {https://link.aps.org/doi/10.1103/PhysRevX.15.011056}
}

@book{nielsen2001quantum,
  title={Quantum computation and quantum information},
  author={Nielsen, Michael A and Chuang, Isaac L},
  year={2001},
  publisher={Cambridge university press Cambridge},
  doi = {https://doi.org/10.1017/CBO9780511976667},
  url = {https://doi.org/10.1017/CBO9780511976667}
}

@article{Tan2021Fisher,
  title = {Fisher Information Universally Identifies Quantum Resources},
  author = {Tan, Kok Chuan and Narasimhachar, Varun and Regula, Bartosz},
  journal = {Physical Review Letters},
  volume = {127},
  issue = {20},
  pages = {200402},
  numpages = {7},
  year = {2021},
  month = {Nov},
  publisher = {American Physical Society},
  doi = {10.1103/PhysRevLett.127.200402},
  url = {https://link.aps.org/doi/10.1103/PhysRevLett.127.200402}
}

@article{Chitambar2019Quantum,
  title = {Quantum resource theories},
  author = {Chitambar, Eric and Gour, Gilad},
  journal = {Reviews of Modern Physics},
  volume = {91},
  issue = {2},
  pages = {025001},
  numpages = {48},
  year = {2019},
  month = {Apr},
  publisher = {American Physical Society},
  doi = {10.1103/RevModPhys.91.025001},
  url = {https://link.aps.org/doi/10.1103/RevModPhys.91.025001}
}

@book{zeng2019quantum,
  title={Quantum information meets quantum matter},
  author={Zeng, Bei and Chen, Xie and Zhou, Duan-Lu and Wen, Xiao-Gang and others},
  year={2019},
  publisher={Springer},
  url = {https://link.springer.com/book/10.1007/978-1-4939-9084-9}
}

@book{bell2004speakable,
  title={Speakable and unspeakable in quantum mechanics: Collected papers on quantum philosophy},
  author={Bell, John Stewart},
  year={2004},
  publisher={Cambridge university press},
  url = {https://doi.org/10.1017/CBO9780511815676} 
}

@article{amico2008entanglement,
 title = {Entanglement in many-body systems},
  author = {Amico, Luigi and Fazio, Rosario and Osterloh, Andreas and Vedral, Vlatko},
  journal = {Reviews of Modern Physics},
  volume = {80},
  issue = {2},
  pages = {517--576},
  numpages = {0},
  year = {2008},
  month = {May},
  publisher = {American Physical Society},
  doi = {10.1103/RevModPhys.80.517},
  url = {https://link.aps.org/doi/10.1103/RevModPhys.80.517}
}

@article{kitaev2006topological,
  title = {Topological Entanglement Entropy},
  author = {Kitaev, Alexei and Preskill, John},
  journal = {Physical Review Letters},
  volume = {96},
  issue = {11},
  pages = {110404},
  numpages = {4},
  year = {2006},
  month = {Mar},
  publisher = {American Physical Society},
  doi = {10.1103/PhysRevLett.96.110404},
  url = {https://link.aps.org/doi/10.1103/PhysRevLett.96.110404}
}

@article{li2008entanglement,
  title = {Entanglement Spectrum as a Generalization of Entanglement Entropy: Identification of Topological Order in Non-Abelian Fractional Quantum Hall Effect States},
  author = {Li, Hui and Haldane, F. D. M.},
  journal = {Physical Review Letters},
  volume = {101},
  issue = {1},
  pages = {010504},
  numpages = {4},
  year = {2008},
  month = {Jul},
  publisher = {American Physical Society},
  doi = {10.1103/PhysRevLett.101.010504},
  url = {https://link.aps.org/doi/10.1103/PhysRevLett.101.010504}
}

@article{Scheie2023Erratum,
  title = {Erratum},
  author = { },
  journal = {Physical Review B},
  volume = {107},
  issue = {5},
  pages = {059902},
  numpages = {2},
  year = {2023},
  month = {Feb},
  publisher = {American Physical Society},
  doi = {10.1103/PhysRevB.107.059902},
  url = {https://link.aps.org/doi/10.1103/PhysRevB.107.059902}
}

@article{Scheie2021Witnessing,
  title = {Witnessing entanglement in quantum magnets using neutron scattering},
  author = {Scheie, A. and Laurell, Pontus and Samarakoon, A. M. and Lake, B. and Nagler, S. E. and Granroth, G. E. and Okamoto, S. and Alvarez, G. and Tennant, D. A.},
  journal = {Physical Review B},
  volume = {103},
  issue = {22},
  pages = {224434},
  numpages = {16},
  year = {2021},
  month = {Jun},
  publisher = {American Physical Society},
  doi = {10.1103/PhysRevB.103.224434},
  url = {https://link.aps.org/doi/10.1103/PhysRevB.103.224434}
}

@Article{scheie2023proximate,
author={Scheie, A. O.
and Ghioldi, E. A.
and Xing, J.
and Paddison, J. A. M.
and Sherman, N. E.
and Dupont, M.
and Sanjeewa, L. D.
and Lee, Sangyun
and Woods, A. J.
and Abernathy, D.
and Pajerowski, D. M.
and Williams, T. J.
and Zhang, Shang-Shun
and Manuel, L. O.
and Trumper, A. E.
and Pemmaraju, C. D.
and Sefat, A. S.
and Parker, D. S.
and Devereaux, T. P.
and Movshovich, R.
and Moore, J. E.
and Batista, C. D.
and Tennant, D. A.},
title={Proximate spin liquid and fractionalization in the triangular antiferromagnet {KYbSe$_2$} },
journal={Nature Physics},
year={2024},
month={Jan},
day={01},
volume={20},
number={1},
pages={74-81},
issn={1745-2481},
doi={10.1038/s41567-023-02259-1},
url={https://doi.org/10.1038/s41567-023-02259-1}
}

@article{Laurell2021Quantifying,
  title = {Quantifying and Controlling Entanglement in the Quantum Magnet ${\mathrm{Cs}}_{2}{\mathrm{CoCl}}_{4}$},
  author = {Laurell, Pontus and Scheie, Allen and Mukherjee, Chiron J. and Koza, Michael M. and Enderle, Mechtild and Tylczynski, Zbigniew and Okamoto, Satoshi and Coldea, Radu and Tennant, D. Alan and Alvarez, Gonzalo},
  journal = {Physical Review Letters},
  volume = {127},
  issue = {3},
  pages = {037201},
  numpages = {7},
  year = {2021},
  month = {Jul},
  publisher = {American Physical Society},
  doi = {10.1103/PhysRevLett.127.037201},
  url = {https://link.aps.org/doi/10.1103/PhysRevLett.127.037201}
}

@article{hales2023witnessing,
	Author = {Hales, Jordyn and Bajpai, Utkarsh and Liu, Tongtong and Baykusheva, Denitsa R. and Li, Mingda and Mitrano, Matteo and Wang, Yao},
	Da = {2023/06/14},
	Doi = {10.1038/s41467-023-38540-3},
	Id = {Hales2023},
	Isbn = {2041-1723},
	Journal = {Nature Communications},
	Number = {1},
	Pages = {3512},
	Title = {Witnessing light-driven entanglement using time-resolved resonant inelastic X-ray scattering},
	Ty = {JOUR},
	Url = {https://doi.org/10.1038/s41467-023-38540-3},
	Volume = {14},
	Year = {2023}}

@article{Baykusheva2023Witnessing,
  title = {Witnessing Nonequilibrium Entanglement Dynamics in a Strongly Correlated Fermionic Chain},
  author = {Baykusheva, Denitsa R. and Kalthoff, Mona H. and Hofmann, Damian and Claassen, Martin and Kennes, Dante M. and Sentef, Michael A. and Mitrano, Matteo},
  journal = {Physical Review Letters},
  volume = {130},
  issue = {10},
  pages = {106902},
  numpages = {8},
  year = {2023},
  month = {Mar},
  publisher = {American Physical Society},
  doi = {10.1103/PhysRevLett.130.106902},
  url = {https://link.aps.org/doi/10.1103/PhysRevLett.130.106902}
}

@ARTICLE{Scheie2024Tutorial,
title = {Tutorial: Extracting entanglement signatures from neutron spectroscopy},
journal = {Materials Today Quantum},
volume = {5},
pages = {100020},
year = {2025},
issn = {2950-2578},
doi = {https://doi.org/10.1016/j.mtquan.2024.100020},
url = {https://www.sciencedirect.com/science/article/pii/S2950257824000209},
author = {Allen Scheie and Pontus Laurell and Wolfgang Simeth and Elbio Dagotto and D. Alan Tennant}
}

@ARTICLE{Mazza2026Quantum,
author={Mazza, Federico
and Biswas, Sounak
and Yan, Xinlin
and Prokofiev, Andrey
and Steffens, Paul
and Si, Qimiao
and Assaad, Fakher F.
and Paschen, Silke},
title={Quantum Fisher information in a strange metal},
journal={Nature Physics},
year={2026},
issn={1745-2481},
url={https://doi.org/10.1038/s41567-026-03298-0}
}

@article{braunstein1994statistical,
  title = {Statistical distance and the geometry of quantum states},
  author = {Braunstein, Samuel L. and Caves, Carlton M.},
  journal = {Physical Review Letters},
  volume = {72},
  issue = {22},
  pages = {3439--3443},
  numpages = {0},
  year = {1994},
  month = {May},
  publisher = {American Physical Society},
  doi = {10.1103/PhysRevLett.72.3439},
  url = {https://link.aps.org/doi/10.1103/PhysRevLett.72.3439}
}

@article{hu2022quantumc,
	author = {Hu, Haoyu and Chen, Lei and Si, Qimiao},
	date = {2024/12/01},
	doi = {10.1038/s41567-024-02679-7},
	id = {Hu2024},
	isbn = {1745-2481},
	journal = {Nature Physics},
	number = {12},
	pages = {1863--1873},
	title = {Quantum critical metals and loss of quasiparticles},
	url = {https://doi.org/10.1038/s41567-024-02679-7},
	volume = {20},
	year = {2024}}

@Article{Si2001,
author={Si, Q.
and Rabello, S.
and Ingersent, K.
and Smith, J. L.},
title={Locally critical quantum phase transitions in strongly correlated metals},
journal={Nature},
year={2001},
month={Oct},
day={01},
volume={413},
number={6858},
pages={804-808},
issn={1476-4687},
doi={10.1038/35101507}
}

@article{Liyang-Chen2023,
author = {Li Yang Chen  and Dale T. Lowder  and Emine Bakali  and Aaron Maxwell Andrews  and Werner Schrenk  and Monika Waas  and Robert Svagera  and Gaku Eguchi  and Lukas Prochaska  and Yiming Wang  and Chandan Setty  and Shouvik Sur  and Qimiao Si  and Silke Paschen  and Douglas Natelson },
title = {Shot noise in a strange metal},
journal = {Science},
volume = {382},
number = {6673},
pages = {907-911},
year = {2023},
doi = {10.1126/science.abq6100},
url = 
{https://www.science.org/doi/epdf/10.1126/science.abq6100}
}

@article{George2020Experimental,
  title = {Experimental realization of multipartite entanglement via quantum Fisher information in a uniform antiferromagnetic quantum spin chain},
  author = {Mathew, George and Silva, Saulo L. L. and Jain, Anil and Mohan, Arya and Adroja, D. T. and Sakai, V. G. and Tomy, C. V. and Banerjee, Alok and Goreti, Rajendar and N., Aswathi V. and Singh, Ranjit and Jaiswal-Nagar, D.},
  journal = {Physical Review Research},
  volume = {2},
  issue = {4},
  pages = {043329},
  numpages = {10},
  year = {2020},
  month = {Dec},
  publisher = {American Physical Society},
  doi = {10.1103/PhysRevResearch.2.043329},
  url = {https://link.aps.org/doi/10.1103/PhysRevResearch.2.043329}
}

@article{Pratt2022Spin,
  title = {Spin dynamics, entanglement, and the nature of the spin liquid state in ${\mathrm{YbZnGaO}}_{4}$},
  author = {Pratt, F. L. and Lang, F. and Steinhardt, W. and Haravifard, S. and Blundell, S. J.},
  journal = {Physical Review B},
  volume = {106},
  issue = {6},
  pages = {L060401},
  numpages = {6},
  year = {2022},
  month = {Aug},
  publisher = {American Physical Society},
  doi = {10.1103/PhysRevB.106.L060401},
  url = {https://link.aps.org/doi/10.1103/PhysRevB.106.L060401}
}

@article{Toth2012Multipartite,
  title = {Multipartite entanglement and high-precision metrology},
  author = {T\'oth, G\'eza},
  journal = {Phys. Rev. A},
  volume = {85},
  issue = {2},
  pages = {022322},
  numpages = {8},
  year = {2012},
  month = {Feb},
  publisher = {American Physical Society},
  doi = {10.1103/PhysRevA.85.022322},
  url = {https://link.aps.org/doi/10.1103/PhysRevA.85.022322}
}

@ARTICLE{Balut2024Quantum,
  title = {Quantum entanglement and quantum geometry measured with inelastic x-ray scattering},
  author = {Ba\l{}ut, David and Bradlyn, Barry and Abbamonte, Peter},
  journal = {Physical Review B},
  volume = {111},
  issue = {12},
  pages = {125161},
  numpages = {7},
  year = {2025},
  month = {Mar},
  publisher = {American Physical Society},
  doi = {10.1103/PhysRevB.111.125161},
  url = {https://link.aps.org/doi/10.1103/PhysRevB.111.125161}
}

@article{Stefan2020,
  title = {Colloquium: Heavy-electron quantum criticality and single-particle spectroscopy},
  author = {Kirchner, Stefan and Paschen, Silke and Chen, Qiuyun and Wirth, Steffen and Feng, Donglai and Thompson, Joe D. and Si, Qimiao},
  journal = {Reviews of Modern Physics},
  volume = {92},
  issue = {1},
  pages = {011002},
  numpages = {19},
  year = {2020},
  month = {Mar},
  publisher = {American Physical Society},
  doi = {10.1103/RevModPhys.92.011002},
  url = {https://link.aps.org/doi/10.1103/RevModPhys.92.011002}
}

@article{Coleman_review,
	Author = {Coleman, Piers and Schofield, Andrew J.},
	Da = {2005/01/01},
	Doi = {10.1038/nature03279},
	Id = {Coleman2005},
	Isbn = {1476-4687},
	Journal = {Nature},
	Number = {7023},
	Pages = {226--229},
	Title = {Quantum criticality},
	Ty = {JOUR},
	Url = {https://doi.org/10.1038/nature03279},
	Volume = {433},
	Year = {2005}}

@article{Coleman2001,
doi = {10.1088/0953-8984/13/35/202},
url = {https://dx.doi.org/10.1088/0953-8984/13/35/202},
year = {2001},
month = {aug},
publisher = {},
volume = {13},
number = {35},
pages = {R723},
author = {P Coleman and  C Pépin and  Qimiao Si and  R Ramazashvili},
title = {How do Fermi
liquids get heavy and die?},
journal = {Journal of Physics: Condensed Matter}
}

@article{Senthil2004,
  title = {Weak magnetism and non-Fermi liquids near heavy-fermion critical points},
  author = {Senthil, T. and Vojta, Matthias and Sachdev, Subir},
  journal = {Physical Review B},
  volume = {69},
  issue = {3},
  pages = {035111},
  numpages = {19},
  year = {2004},
  month = {Jan},
  publisher = {American Physical Society},
  doi = {10.1103/PhysRevB.69.035111},
  url = {https://link.aps.org/doi/10.1103/PhysRevB.69.035111}
}

@INPROCEEDINGS{petz2011introduction,
       author = {{Petz}, D. and {Ghinea}, C.},
        title = "{Introduction to Quantum Fisher Information}",
    booktitle = {Quantum Probability and Related Topics},
         year = 2011,
       editor = {{Rebolledo}, Rolando and {Orszag}, Miguel},
        month = jan,
        pages = {261-281},
          doi = {10.1142/9789814338745_0015},
url = {https://doi.org/10.1142/9789814338745_0015},
archivePrefix = {arXiv},
       eprint = {1008.2417},
 primaryClass = {quant-ph},
       adsurl = {https://ui.adsabs.harvard.edu/abs/2011qprt.conf..261P},
}

@article{paschen2021quantum,
	Author = {Paschen, Silke and Si, Qimiao},
	Da = {2021/01/01},
	Doi = {10.1038/s42254-020-00262-6},
	Id = {Paschen2021},
	Isbn = {2522-5820},
	Journal = {Nature Reviews Physics},
	Number = {1},
	Pages = {9--26},
	Title = {Quantum phases driven by strong correlations},
	Ty = {JOUR},
	Url = {https://doi.org/10.1038/s42254-020-00262-6},
	Volume = {3},
	Year = {2021}}

@article{hu2022extended,
       author = {{Hu}, Haoyu and {Chen}, Lei and {Si}, Qimiao},
        title = "{Extended Dynamical Mean Field Theory for Correlated Electron Models}",
      journal = {arXiv e-prints},
         year = 2022,
        month = oct,
          eid = {arXiv:2210.14197},
        pages = {arXiv:2210.14197},
          doi = {10.48550/arXiv.2210.14197},
url = {https://doi.org/10.48550/arXiv.2210.14197},
archivePrefix = {arXiv},
       eprint = {2210.14197},
 primaryClass = {cond-mat.str-el},
       adsurl = {https://ui.adsabs.harvard.edu/abs/2022arXiv221014197H}
}

@article{Si1996,
  title = {Kosterlitz-Thouless Transition and Short Range Spatial Correlations in an Extended Hubbard Model},
  author = {Si, Qimiao and Smith, J. Lleweilun},
  journal = {Physical Review Letters},
  volume = {77},
  issue = {16},
  pages = {3391--3394},
  numpages = {0},
  year = {1996},
  month = {Oct},
  publisher = {American Physical Society},
  doi = {10.1103/PhysRevLett.77.3391},
  url = {https://link.aps.org/doi/10.1103/PhysRevLett.77.3391}
}

@article{Smith2000,
  title = {Spatial correlations in dynamical mean-field theory},
  author = {Smith, J. Lleweilun and Si, Qimiao},
  journal = {Physical Review B},
  volume = {61},
  issue = {8},
  pages = {5184--5193},
  numpages = {0},
  year = {2000},
  month = {Feb},
  publisher = {American Physical Society},
  doi = {10.1103/PhysRevB.61.5184},
  url = {https://link.aps.org/doi/10.1103/PhysRevB.61.5184}
}

@article{Chitra2000,
  title = {Effect of Long Range Coulomb Interactions on the Mott Transition},
  author = {Chitra, R. and Kotliar, G.},
  journal = {Physical Review Letters},
  volume = {84},
  issue = {16},
  pages = {3678--3681},
  numpages = {0},
  year = {2000},
  month = {Apr},
  publisher = {American Physical Society},
  doi = {10.1103/PhysRevLett.84.3678},
  url = {https://link.aps.org/doi/10.1103/PhysRevLett.84.3678}
}

@article{Scheie2024Reconstructing,
  title = {Reconstructing the spatial structure of quantum correlations in materials},
  author = {Scheie, Allen and Laurell, Pontus and Dagotto, Elbio and Tennant, D. Alan and Roscilde, Tommaso},
  journal = {Physical Review Research},
  volume = {6},
  issue = {3},
  pages = {033183},
  numpages = {11},
  year = {2024},
  month = {Aug},
  publisher = {American Physical Society},
  doi = {10.1103/PhysRevResearch.6.033183},
  url = {https://link.aps.org/doi/10.1103/PhysRevResearch.6.033183}
}

@ARTICLE{wang2025local,
       author = {{Wang}, Yiming and {Fang}, Yuan and {Xie}, Fang and {Si}, Qimiao},
        title = "{Local and Non-local Entanglement Witnesses of Fermi Liquid}",
      journal = {arXiv e-prints},
         year = 2025,
        month = feb,
          eid = {arXiv:2502.13958},
        pages = {arXiv:2502.13958},
          doi = {10.48550/arXiv.2502.13958},
archivePrefix = {arXiv},
       eprint = {2502.13958},
 primaryClass = {cond-mat.str-el},
       adsurl = {https://ui.adsabs.harvard.edu/abs/2025arXiv250213958W}
}

@article{BALUT20251354750,
title = {Quantum fisher information reveals UV-IR mixing in the strange metal},
journal = {Physica C: Superconductivity and its Applications},
volume = {635},
pages = {1354750},
year = {2025},
issn = {0921-4534},
doi = {https://doi.org/10.1016/j.physc.2025.1354750},
url = {https://www.sciencedirect.com/science/article/pii/S0921453425001030},
author = {David Bałut and Xuefei Guo and Niels de Vries and Dipanjan Chaudhuri and Barry Bradlyn and Peter Abbamonte and Philip W. Phillips}
}

@ARTICLE{Ji2025Density,
       author = {{Ji}, Guangyue and {Palomino}, David E. and {Goldman}, Nathan and {Ozawa}, Tomoki and {Riseborough}, Peter and {Wang}, Jie and {Mera}, Bruno},
        title = "{Density Matrix Geometry and Sum Rules}",
      journal = {arXiv e-prints},
         year = 2025,
        month = jul,
          eid = {arXiv:2507.14028},
        pages = {arXiv:2507.14028},
          doi = {10.48550/arXiv.2507.14028},
archivePrefix = {arXiv},
       eprint = {2507.14028},
 primaryClass = {cond-mat.mes-hall},
       adsurl = {https://ui.adsabs.harvard.edu/abs/2025arXiv250714028J}
}

@ARTICLE{Guan2025Exploring,
  title = {Exploring many-body quantum geometry beyond the quantum metric with correlation functions: A time-dependent perspective},
  author = {Guan, Yuntao and Bradlyn, Barry},
  journal = {Physical Review Research},
  volume = {8},
  issue = {1},
  pages = {013291},
  numpages = {28},
  year = {2026},
  month = {Mar},
  publisher = {American Physical Society},
  doi = {10.1103/3xjs-c7v7},
  url = {https://link.aps.org/doi/10.1103/3xjs-c7v7}
}

@article{MerminWagner,
  title = {Absence of Ferromagnetism or Antiferromagnetism in One- or Two-Dimensional Isotropic Heisenberg Models},
  author = {Mermin, N. D. and Wagner, H.},
  journal = {Physical Review Letters},
  volume = {17},
  issue = {22},
  pages = {1133--1136},
  numpages = {0},
  year = {1966},
  month = {Nov},
  publisher = {American Physical Society},
  doi = {10.1103/PhysRevLett.17.1133},
  url = {https://link.aps.org/doi/10.1103/PhysRevLett.17.1133}
}

@article{manousakis1991rmp,
  title = {The spin-\textonehalf{} Heisenberg antiferromagnet on a square lattice and its application to the cuprous oxides},
  author = {Manousakis, Efstratios},
  journal = {Reviews of Modern Physics},
  volume = {63},
  issue = {1},
  pages = {1--62},
  numpages = {0},
  year = {1991},
  month = {Jan},
  publisher = {American Physical Society},
  doi = {10.1103/RevModPhys.63.1},
  url = {https://link.aps.org/doi/10.1103/RevModPhys.63.1}
}

@book{auerbach2012interacting,
  title={Interacting electrons and quantum magnetism},
  author={Auerbach, Assa},
  year={2012},
  publisher={Springer Science \& Business Media}
}

@article{KAUFMANN2023108519,
title = {ana_cont: Python package for analytic continuation},
journal = {Computer Physics Communications},
volume = {282},
pages = {108519},
year = {2023},
issn = {0010-4655},
doi = {https://doi.org/10.1016/j.cpc.2022.108519},
url = {https://www.sciencedirect.com/science/article/pii/S0010465522002387},
author = {Josef Kaufmann and Karsten Held}
}

@article{Frerot2016QV,
  title = {Quantum variance: A measure of quantum coherence and quantum correlations for many-body systems},
  author = {Fr\'erot, Ir\'en\'ee and Roscilde, Tommaso},
  journal = {Physical Review B},
  volume = {94},
  issue = {7},
  pages = {075121},
  numpages = {15},
  year = {2016},
  month = {Aug},
  publisher = {American Physical Society},
  doi = {10.1103/PhysRevB.94.075121},
  url = {https://link.aps.org/doi/10.1103/PhysRevB.94.075121}
}

@article{Laurell2024Witnessing,
author = {Laurell, Pontus and Scheie, Allen and Dagotto, Elbio and Tennant, D. Alan},
title = {Witnessing Entanglement and Quantum Correlations in Condensed Matter: A Review},
journal = {Advanced Quantum Technologies},
volume = {8},
number = {3},
pages = {2400196},
doi = {https://doi.org/10.1002/qute.202400196},
url = {https://advanced.onlinelibrary.wiley.com/doi/abs/10.1002/qute.202400196},
year = {2025}
}

@article{Bippus2025Entanglement,
  title = {Entanglement in the pseudogap regime of cuprate superconductors},
  author = {Bippus, Frederic and Krsnik, Juraj and Kitatani, Motoharu and Ak\ifmmode \check{s}\else \v{s}\fi{}amovi\ifmmode \acute{c}\else \'{c}\fi{}, Luka and Kauch, Anna and Bari\ifmmode \check{s}\else \v{s}\fi{}i\ifmmode \acute{c}\else \'{c}\fi{}, Neven and Held, Karsten},
  journal = {Physical Review B},
  volume = {112},
  issue = {8},
  pages = {L081110},
  numpages = {7},
  year = {2025},
  month = {Aug},
  publisher = {American Physical Society},
  doi = {10.1103/xk42-b9cx},
  url = {https://link.aps.org/doi/10.1103/xk42-b9cx}
}

@article{Frerot2019QV,
	author = {Fr{\'e}rot, Ir{\'e}n{\'e}e and Roscilde, Tommaso},
	date = {2019/02/04},
	doi = {10.1038/s41467-019-08324-9},
	id = {Fr{\'e}rot2019},
	isbn = {2041-1723},
	journal = {Nature Communications},
	number = {1},
	pages = {577},
	title = {Reconstructing the quantum critical fan of strongly correlated systems using quantum correlations},
	url = {https://doi.org/10.1038/s41467-019-08324-9},
	volume = {10},
	year = {2019}}

@article{frerot2024symmetry,
  title = {Symmetry: A Fundamental Resource for Quantum Coherence and Metrology},
  author = {Fr\'erot, Ir\'en\'ee and Roscilde, Tommaso},
  journal = {Physical Review Letters},
  volume = {133},
  issue = {26},
  pages = {260402},
  numpages = {6},
  year = {2024},
  month = {Dec},
  publisher = {American Physical Society},
  doi = {10.1103/PhysRevLett.133.260402},
  url = {https://link.aps.org/doi/10.1103/PhysRevLett.133.260402}
}

\onecolumngrid
\begin{center}
\textbf{\large End Matter}
\end{center}
\twocolumngrid
\blue{\emph{Brief review of quantum Fisher information}}---
QFI was originally introduced in quantum metrology as a measure of the sensitivity of a quantum state to changes in a parameter~\cite{braunstein1994statistical}.
As an entanglement witness, the QFI is defined for a quantum state $\rho$ and a Hermitian operator $\widehat{O}$ by
$F_Q = 2\sum_{ij}\frac{(\lambda_i-\lambda_j)^2}{\lambda_i+\lambda_j} {O}_{ij} {O}_{ji}$,
where $\lambda_i$ and $|i\rangle$ are the eigenvalues and eigenstates of $\rho$, ${O}_{ij} = \langle i|\widehat{O}|j\rangle$ and the sum runs over all pairs of $\lambda_i+\lambda_j> 0$~\cite{braunstein1994statistical,petz2011introduction,Liu2020Quantum}.
For a system of $N$ sites, it is convenient to define the QFI density $f_Q = F_Q/N$.
A quantum state has entanglement depth $m$ (also called $m$-producible) if its density matrix can be written as a convex mixture of product states on divided subsystems, $\rho = \sum_{\alpha} p_\alpha \otimes_{\ell} \rho_{\alpha,\ell}$ where $\sum_{\alpha} p_\alpha=1$, $p_\alpha\geq 0$ and each subsystem $\ell$ contains at most $m$ sites.
States that are $m$-producible but not $m-1$-producible are said $m$-partite entangled.
For an $m$-producible state and an operator $\widehat{O}=\sum_j \widehat{O}_j$ built from local terms whose single site spectrum lies in $[h_{\mathrm{min}},h_{\mathrm{max}}]$, the QFI density is bounded above by $m\,(h_{\mathrm{max}}-h_{\mathrm{min}})^2$.
In particular, if~\cite{hyllus2012fisher,Toth2012Multipartite}
\begin{equation}
    f_Q > m\,(h_{\mathrm{max}}-h_{\mathrm{min}})^2 \, ,
\end{equation}
then the QFI witnesses an entanglement depth of at least $m+1$ in the state. The bound is valid as long as each local operator is Hermitian up to a constant phase factor (a proof can be found in Ref.~\cite{wang2025local}).
The normalized QFI density is defined as nQFI$=f_Q/(h_{\mathrm{max}}-h_{\mathrm{min}})^2$. 
For spin-$1/2$ operators $\widehat{S^\alpha}$ where $\alpha=x,y,z$, nQFI coincides with $f_Q$. 
In the following, we focus on this case and use $f_Q^\alpha$ to denote the normalized QFI density of local spin operators $\widehat{O}_j = \widehat{S_j^{\alpha}}$ ($\alpha=x,y,z$) with site-dependent phase factors.

When $\widehat{O}_{\mathbf{q}} = \sum_{j=1}^N  e^{i\mathbf{q}\cdot \mathbf{r}_j} \widehat{O_j}$, where $\widehat{O_j}$ is local Hermitian operator, the QFI density can be expressed in terms of the dynamical susceptibility as~\cite{hauke2016measuring}
\begin{equation}
f_{Q}(\mathbf{q}) = \frac{4}{\pi} \int_{0}^{\infty} d\nu  \tanh \frac{\beta \nu}{2}  \chi^{\prime\prime}(\mathbf{q},\nu) \,,
\end{equation}
where $\chi^{\prime\prime}(\mathbf{q},\nu)$ denotes the imaginary part of the retarded susceptibility of the operator $\widehat{O_{\mathbf{q}}}$.
This establishes a direct connection between QFI and experimentally accessible dynamical response functions.

For numerical evaluations, it is useful to rewrite the QFI density as
\begin{equation}
    f_{Q}(\mathbf{q}) = 4 S(\mathbf{q}) - \frac{4}{\pi} \int_{0}^{\infty} d\nu  \frac{2}{\sinh(\beta \nu)}  \chi^{\prime\prime}(\mathbf{q},\nu) \,,
\label{eqn:qfi_split}
\end{equation}
where $S(\mathbf{q})$ is the static structure factor.
At low temperature Eq.~(\ref{eqn:qfi_split}) reduces to the equal-time formula $f_{Q}(\mathbf{q}) \approx 4 S(\mathbf{q})$.
Eq.~(\ref{eqn:qfi_split}) separates equal-time correlations from a finite frequency correction term, and it is well suited for numerical calculations based on imaginary time data such as EDMFT.
The static structure factor can be obtained by $S(\mathbf{q})=\chi(\mathbf{q},\tau=0^-)$ and the real-frequency dynamical susceptibility $\chi^{\prime\prime}(\mathbf{q},\nu)$ is obtained via analytic continuation of the imaginary time data, using methods such as Pad\'{e} approximation and maximum entropy~\cite{KAUFMANN2023108519}.
In practice, these continuation methods tend to agree on the low frequency response, which provides the dominant contribution to the correction term in Eq.~(\ref{eqn:qfi_split}) (The integrand is finite at $\nu\to 0$ where $\chi(\mathbf{q},\nu)\propto \nu$).
In the maintext, Pad\'{e} method is used for all the EDMFT results, except at the QCP where the maximum entropy method is used.

Another relevant quantity is quantum variance (QV), which is defined as the difference between total variance and thermal fluctuation induced variance.
It provides a lower bound for QFI: QV$\leq f_Q$~\cite{Scheie2024Reconstructing,Laurell2024Witnessing,Bippus2025Entanglement}. 
We leave the detailed analysis on QV in SM~\cite{sm}.

\blue{\emph{Linear spin wave theory solutions}}---
We use Holstein--Primakoff transformation to analyze the N\'{e}el AF-ordered phase.
In the long-wavelength limit, the spin wave (Goldstone mode) has a linear dispersion
\begin{align}
    \omega_{\mathbf{q}} = z \alpha J_1 \bar{S} \sqrt{1 - \gamma_{\mathbf{q}}^2} \approx \sqrt{2z} \alpha J_1 \bar{S} |\widetilde{\mathbf{q}}|
\end{align}
where $\mathbf{q}$ is the momentum away from $\mathbf{Q}$, $|\widetilde{\mathbf{q}}|=\sqrt{q_\parallel^2+\lambda q_z^2}$, $\bar S = (S+m_z)/2$ and $m_z$ is the ordered moment, which can be solved self-consistently~\cite{manousakis1991rmp,sm}.
Here $z=4+2\lambda$ is the effective coordination number and $\gamma_{\mathbf{q}} = \frac{1}{z} \left[2\cos q_x + 2\cos q_y + 2 \lambda \cos q_z\right]$ for our tetragonal lattice.
The spin wave velocity is renormalized by the frustrating interaction $J_2$ through a factor $\alpha\approx\sqrt{1-2 J_2/J_1}$ which vanishes in the limit $J_2/J_1 \to 1/2$.
This softening of the Goldstone mode signals the instability of N\'{e}el order and enhances low-energy quantum fluctuations.

\clearpage
\onecolumngrid
\appendix
\begin{center}
\textbf{\large \titlename 
\vspace{4pt} \\ 
SUPPLEMENTAL MATERIAL}
\end{center}

\section*{Analytic continuation}\label{sec:analytic_continuation}
\begin{figure}[th]
    \centering
    \includegraphics[width=0.5\linewidth]{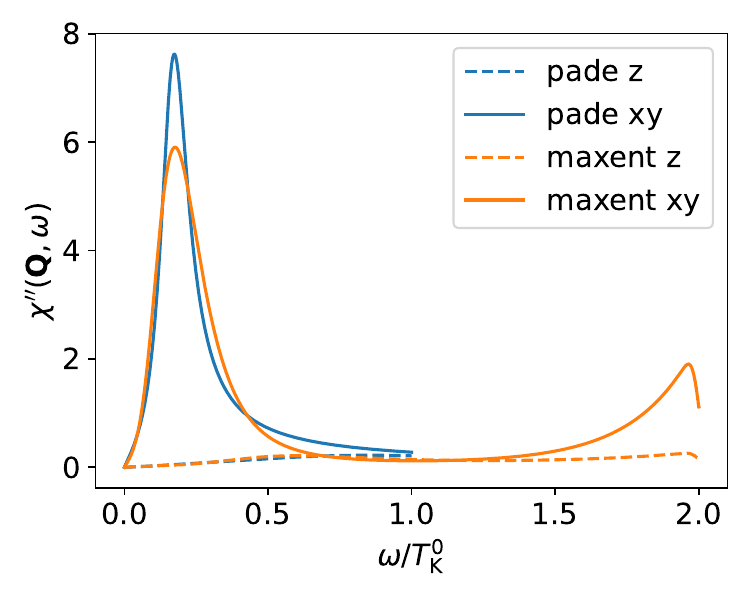}
    \caption{Analytic continued $\chi''(\mathbf{Q},\omega)$ using Pad\'{e} and maximum entropy for longitudinal and transverse spin operators. 
    Here $\delta=I/T_{\mathrm{K}}^0=1.0$ and $T/T_{\mathrm{K}}^0=0.01$. 
    Both methods generate similar $\chi''$ at low frequency $\omega \sim T$. Although maximum entropy has artificial features at large frequency $\omega \gg T$ does not affect our QFI calculation.
    }
    \label{fig:ana_cont}
\end{figure}

We solved imaginary frequency dynamic spin susceptibilities using extended dynamical mean-field theory (EDMFT) for both longitudinal $\widehat{S^z}$ and transverse $\widehat{S^x}$(or $\widehat{S^x}$) components. 
In order to compute quantum Fisher information (QFI), we need to perform analytic continuation to the spin susceptibilities.
We used Pad\'{e} approximation and maximum entropy methods ~\cite{KAUFMANN2023108519}. 
The two methods generate qualitatively compatible results. We present comparison of the two methods with one example in Fig.~\ref{fig:ana_cont}. 
Here $\delta=I/T_{\mathrm{K}}^0=1.0$ and $T/T_{\mathrm{K}}^0=0.01$. 

\begin{figure}[th]
    \centering
    \includegraphics[width=0.5\linewidth]{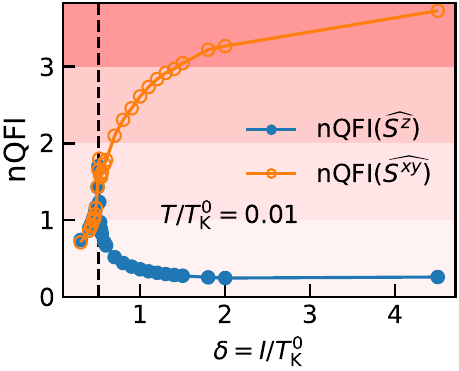}
    \caption{Normalized QFI density determined through maximum entropy method. Only converged points are shown here.}
    \label{fig:nQFI_ME}
\end{figure}

The normalized QFI density (nQFI) determined from maximum entropy is shown in Fig.~\ref{fig:nQFI_ME} for comparison with the nQFI determined from Pad\'{e} method that is shown in Fig.~2(a) in maintext.

\section*{Quantum Variance}\label{sec:quantum_variance}
Quantum variance is an entanglement witness that is defined by the total variance subtracting thermal fluctuation induced variance~\cite{Frerot2016QV,Frerot2019QV}. 
For a thermal state in the equilibrium, it is determined by~\cite{Scheie2024Reconstructing,Laurell2024Witnessing}
\begin{equation}
    I_Q(\widehat{O}) = \frac{1}{\pi} \int_0^\infty d\nu \left(\coth\frac{\beta\nu}{2}-\frac{2}{\beta\nu}\right) \chi_{\widehat{O}}^{\prime\prime}(\nu) \,,
\end{equation}
where $\chi_{\widehat{O}}^{\prime\prime}(\nu) $ is the imaginary part of the retarded susceptibility for Hermitian operator $\widehat{O}$. 
Using identities $\tanh \frac{z}{2} = \sum_{n=-\infty}^\infty 2z /((\pi (2n+1))^2+z^2)$ and $\coth \frac{z}{2} = \sum_{n=-\infty}^\infty 2z /((2\pi n)^2+z^2)$, we have 
\begin{align}
    4 I_Q \leq f_Q \,.
\end{align}
Note this inequality is valid for generic mixed states\cite{Frerot2016QV}.
QV can be determined solely from imaginary frequency susceptibility by~\cite{Bippus2025Entanglement}
\begin{equation}
    I_Q = \frac{2}{\beta}\sum_{n=1}^\infty \chi_{\widehat{O}}^{\prime}(i\nu_n) \,,
\end{equation}
where $\chi_{\widehat{O}}(i\nu_n)$ is the real part of the Matsubara susceptibility (for spin operators Matsubara susceptibility is always real), and $i\nu_n = 2\pi n/\beta$, $n\in \mathbb Z$.
This expression explicitly shows that the zero frequency information is lost for QV, which makes it not a good lower bound for QFI at quantum critical point where the zero frequency contribution is significant.
For states in the ordered phases, QV could be good approximation for QFI.

\begin{figure}[ht]
    \centering
    \includegraphics[width=0.5\linewidth]{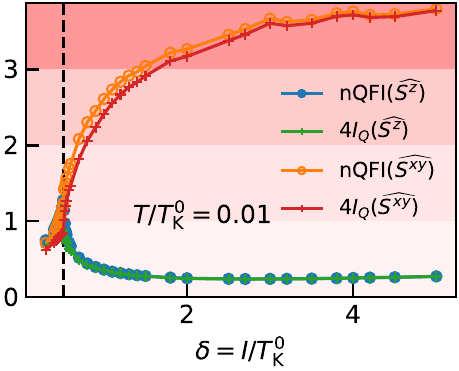}
    \caption{Comparison of nQFI (using Pad\'{e}) and QV for the Kondo lattice model at $T/T_{\mathrm{K}}^0=0.01$.}
    \label{fig:QFI_QV}
\end{figure}

The longitudinal and transverse nQFI and QV for the Kondo lattice model at $T/T_{\mathrm{K}}^0=0.01$ are shown in Fig.~\ref{fig:QFI_QV}. nQFI and QV show good agreement at large $\delta$.

\section*{Spin wave analysis for Heisenberg model}\label{sec:spin_wave_heisenberg}
Holstein--Primakoff transformation for AF spin wave is given by:
\begin{equation}
    \begin{aligned}
        \widehat{S_{i,A}^{+}} &= \sqrt{2S-a_i^\dagger a_i} a_{i}, \quad \widehat{S_{i,A}^{-}} = a_{i}^{\dagger}\sqrt{2S-a_i^\dagger a_i} , \quad \widehat{S_{i,A}^{z}} = S - a_{i}^{\dagger} a_{i}, \\
        \widehat{S_{i,B}^{+}} &=  b_{i}^{\dagger} \sqrt{2S-b_i^\dagger b_i} , \quad \widehat{S_{i,B}^{-}} = \sqrt{2S-b_i^\dagger b_i} b_{i}, \quad \widehat{S_{i,B}^{z}} = -S + b_{i}^{\dagger} b_{i}.
    \end{aligned}
\end{equation}
where $A/B$ is the sublattice label.
We approximate ${2S - a_i^\dagger a_i} \approx {S+m_z}$, where $m_z$ is the ordered moment. Then the transformation reduces to:
\begin{equation}
    \begin{aligned}
        \widehat{S_{i,A}^{+}} &= \sqrt{S+m_z} a_{i}, \quad \widehat{S_{i,A}^{-}} = \sqrt{S+m_z} a_{i}^{\dagger}, \quad \widehat{S_{i,A}^{z}} = S - a_{i}^{\dagger} a_{i}, \\
        \widehat{S_{i,B}^{+}} &= \sqrt{S+m_z} b_{i}^{\dagger}, \quad \widehat{S_{i,B}^{-}} = \sqrt{S+m_z} b_{i}, \quad \widehat{S_{i,B}^{z}} = -S + b_{i}^{\dagger} b_{i}.
    \end{aligned}
\end{equation}
Then the Heisenberg Hamiltonian for the AF-ordered phase can be written as:
\begin{align}
    H &= \sum_{\langle i,j \rangle} J_{ij} \left[
    \frac{1}{2}\left( S_{i,A}^{+} S_{j,B}^{-} + S_{i,A}^{-} S_{j,B}^{+} \right)
    + S_{i,A}^{z} S_{j,B}^{z} \right] \\ 
    &= \sum_{\langle i,j\rangle} J_{ij}
    \left[ \bar{S}\left(a_i b_j + a_i^\dagger b_j^\dagger
     + a_i^\dagger a_i + b_j^\dagger b_j \right)
    - a_i^\dagger a_i b_j^\dagger b_j - S^2 \right] \\
    &\approx \sum_{\langle i,j\rangle} J_{ij} \bar{S} \left(
    a_i b_j + a_i^\dagger b_j^\dagger + a_i^\dagger a_i + b_j^\dagger b_j \right) 
\end{align} 
where $\bar{S}=(S+m_z)/2$, the interaction term is negligible in large $S$ limit and constant term is ignored.
We consider $J_{ij}=J$ for in-plane neighbors and $J_{ij}=\lambda J$ for out-of-plane neighbors on a tetragonal lattice.
To diagonalize the Hamiltonian, we first perform the Fourier transform:
\begin{align}
    a_{i} &= \frac{1}{\sqrt{N/2}}\sum_{\mathbf{k}} a_{\mathbf{k}} e^{-i \mathbf{k} \cdot \mathbf{r}_i}\,, \quad a_{\mathbf{k}} = \frac{1}{\sqrt{N/2}} \sum_i a_{i} e^{i \mathbf{k} \cdot \mathbf{r}_i} \\
    b_{i} &= \frac{1}{\sqrt{N/2}}\sum_{\mathbf{k}} b_{\mathbf{k}} e^{-i \mathbf{k} \cdot \mathbf{r}_j}\,, \quad b_{\mathbf{k}} = \frac{1}{\sqrt{N/2}} \sum_i b_{i} e^{i \mathbf{k} \cdot \mathbf{r}_i} 
\end{align}
where $N$ is the total number of sites.
Substituting the Fourier transforms into the Hamiltonian, we get:
\begin{align}
    H &\approx zJ \bar{S} \sum_{\mathbf{k}}
    \gamma_{\mathbf{k}} a_{\mathbf{k}} b_{-\mathbf{k}} + \gamma_{\mathbf{k}}^* a_{\mathbf{k}}^\dagger b_{-\mathbf{k}}^\dagger
     + a_{\mathbf{k}}^\dagger a_{\mathbf{k}} + b_{-\mathbf{k}}^\dagger b_{-\mathbf{k}} 
\end{align}
where $z=4+2\lambda$ is the effective coordination number of the lattice and $\gamma_{\mathbf{k}} = \frac{1}{z} \left[2\cos k_x + 2\cos k_y + 2 \lambda \cos k_z\right]$ for our tetragonal lattice.

Apply the Bogoliubov transformation to diagonalize the Hamiltonian, we define new bosonic operators $\alpha_{\mathbf{k}}$ and $\beta_{\mathbf{k}}$ as follows:
\begin{align}
    a_{\mathbf{k}} &= u_{\mathbf{k}} \alpha_{\mathbf{k}} + v_{\mathbf{k}} \beta_{-\mathbf{k}}^\dagger \,, \\
    b_{-\mathbf{k}}^\dagger &= u_{\mathbf{k}} \beta_{-\mathbf{k}}^\dagger + v_{\mathbf{k}} \alpha_{\mathbf{k}} \,.
\end{align}
where $u_{\mathbf{k}}$ and $v_{\mathbf{k}}$ are:
\begin{align}
    u_{\mathbf{k}} &= \sqrt{\frac{1}{2} \left( \frac{z J \bar{S}}{\omega_{\mathbf{k}}} + 1 \right)} \,, \\
    v_{\mathbf{k}} &= -\sqrt{\frac{1}{2} \left( \frac{z J \bar{S}}{\omega_{\mathbf{k}}} - 1 \right)} \,.
\end{align}
The diagonalized Hamiltonian can be written as:
\begin{align}
    H = \sum_{\mathbf{k}} \omega_{\mathbf{k}} \left( \alpha_{\mathbf{k}}^\dagger \alpha_{\mathbf{k}} + \beta_{-\mathbf{k}}^\dagger \beta_{-\mathbf{k}} \right) + \text{const.} \,,
\end{align}
and the spin wave dispersion is:
\begin{align}
    \omega_{\mathbf{k}} = z J \bar{S} \sqrt{1 - \gamma_{\mathbf{k}}^2} \approx \sqrt{2z}J \bar{S} |\widetilde{\mathbf{k}}|
\end{align}
where $|\widetilde{\mathbf{k}}|=\sqrt{k_\parallel^2+\lambda k_z^2}$.
Using this notation, we can express $u_{\mathbf{k}}v_{\mathbf{k}}=-\frac12 \frac{\gamma_k}{\sqrt{1-\gamma_k^2}}$ and $u_{\mathbf{k}}^2 + v_{\mathbf{k}}^2 = \frac{1}{\sqrt{1-\gamma_k^2}}$.


Note the ground state is annihilated by Bogoliubov operators $\alpha_{\mathbf k} |\varnothing \rangle = \beta_{\mathbf k} |\varnothing \rangle = 0$ for all $\mathbf k$, which is a highly entangled state in the original basis.

\subsection*{Transverse spin QFI}
Define operator for QFI:
\begin{equation}
    \widehat{O} = \sum_{i=1}^N \phi_i \widehat{S^x_i} = \sum_{i=1}^N \phi_i \frac{\widehat{S^+_i} + \widehat{S^-_i}}{2}
\end{equation}
where $\phi_j=\exp(i {\mathbf K}\cdot \mathbf{R}_j)$. 
Define the momentum away from ordering vector ${\mathbf k}={\mathbf K} - {\mathbf Q}$.
The normalized QFI density is
\begin{align}
    \mathrm{nQFI}(\widehat{S^{xy}},{\mathbf k},T) &= \frac{1}{4S^2} \frac{4}{\pi} \int_{0}^{\infty} d\omega \tanh\left( \frac{\beta \omega}{2} \right) \text{Im} \chi^{xx}(\mathbf{k}, \omega) \\
    &= \frac{1}{4S^2} \frac{4}{\pi} \int_{0}^{\infty} d\omega \tanh\left( \frac{\beta \omega}{2} \right) \text{Im} \frac{\chi^{+-}(\mathbf{k}, \omega) + \chi^{-+}(\mathbf{k}, \omega) }{4}
\end{align}
Here
\begin{align}
    \widehat{S_{\mathbf{k}}^+} &= \frac{1}{\sqrt{N}}\sum_{j=1}^N e^{i({\mathbf k} + {\mathbf Q})\cdot {\mathbf R}_j}S_j^+ \\
    &= \frac{1}{\sqrt{N}}\sum_{j=1}^{N/2} e^{i{\mathbf k}\cdot {\mathbf r}_j} (S_{j,A}^+ - S_{j,B}^+) \\ 
    &= \frac{1}{\sqrt{2}} \sqrt{2 \bar{S}} (a_{\mathbf{k}} - b_{-\mathbf{k}}^\dagger)
\end{align}
We have assumed $|\mathbf{k}|\ll 1$ to simplify $e^{i\mathbf{k}\cdot \mathbf{\delta}}=1$.
Similarly, we can express all the spin operators in terms of the Bogoliubov quasiparticles:
\begin{align}
    \widehat{S^+_{\mathbf{k}}} &= \sqrt{\bar{S}} (a_{\mathbf{k}} - b_{-\mathbf{k}}^\dagger) 
    = \sqrt{\bar{S}} (u_{\mathbf{k}} - v_{\mathbf{k}}) (\alpha_{\mathbf{k}} -  \beta_{-\mathbf{k}}^\dagger )\\ 
    \widehat{S^-_{\mathbf{k}}} &= \sqrt{\bar{S}} (a_{\mathbf{k}}^\dagger - b_{-\mathbf{k}})
    = \sqrt{\bar{S}} (u_{\mathbf{k}} - v_{\mathbf{k}}) (\alpha_{\mathbf{k}}^\dagger -  \beta_{-\mathbf{k}} ) 
\end{align}

Use the retarded bosonic Green's function
\begin{align}
    G^R(\omega)&= -i\int dt e^{i\omega t} \theta(t)\langle [\alpha_{\mathbf{k}}(t),\alpha_{\mathbf{k}}^\dagger(0)] \rangle \\ 
    &= \frac{1}{\omega - \omega_{\mathbf{k}} + 0^+}
\end{align}
Then the retarded spin susceptibilities are:
\begin{align}
    \chi^{+-}(\mathbf{k}, \omega) = 
    \chi^{-+}(\mathbf{k},\omega) &= \bar{S} \sqrt{\frac{1 + \gamma_{\mathbf{k}}}{1 - \gamma_{\mathbf{k}}}}
    \left( \frac{1}{\omega - \omega_{\mathbf{k}} + i0^+} - \frac{1}{\omega + \omega_{\mathbf{k}} + i0^+} \right)
\end{align}
Therefore, the normalized QFI density for alternating $\widehat{S^{x}}$ or $\widehat{S^{y}}$ is:
\begin{align}
    \mathrm{nQFI}(\widehat{S^{xy}},{\mathbf k},T) &= \frac{1}{4S^2} \frac{4}{\pi} \int_{0}^{\infty} d\omega \tanh\left( \frac{\beta \omega}{2} \right) \text{Im} \frac{\chi^{+-}(\mathbf{k}, \omega)}{2} \\
    &= \frac{1}{4S^2} 2\bar{S} \sqrt{\frac{1 + \gamma_{\mathbf{k}}}{1 - \gamma_{\mathbf{k}}}} \tanh\left( \frac{\beta \omega_{\mathbf{k}}}{2} \right) \\
    &\approx \frac{z J_1}{\omega_{\mathbf q}} 
    \left(\frac{\bar S}{S}\right)^2 \tanh\left( \frac{\beta \omega_{\mathbf q}}{2} \right) \\ 
    &\approx \frac{1}{4S^2} \frac{4\sqrt{2+\lambda}\bar{S}}{|\widetilde{\mathbf{k}}|} \tanh\left( \sqrt{2+\lambda}\beta J \bar{S} |\widetilde{\mathbf{k}}| \right) 
\end{align}
At $\mathbf{k} = 0$, the normalized QFI density is:
\begin{equation}
    \mathrm{nQFI}(\widehat{S^{xy}},{\mathbf k} = 0,T) = (2+\lambda) \beta J \left(\frac{\bar S}{S}\right)^2 
\end{equation}
At $T=0$, the normalized QFI density is:
\begin{equation}
    \mathrm{nQFI}(\widehat{S^{xy}},{\mathbf k},T=0) = \frac{\sqrt{2+\lambda}\bar{S}}{S^2 |\widetilde{\mathbf{k}}|}
\end{equation}
At $S\to \infty$ limit, the normalized QFI density is:
\begin{equation}
    \lim_{S\to \infty} \mathrm{nQFI}(\widehat{S^{xy}},{\mathbf k},T) = 0
\end{equation}

\subsection*{Longitudinal spin QFI}
\begin{figure}
    \centering
    \includegraphics[width=0.5\linewidth]{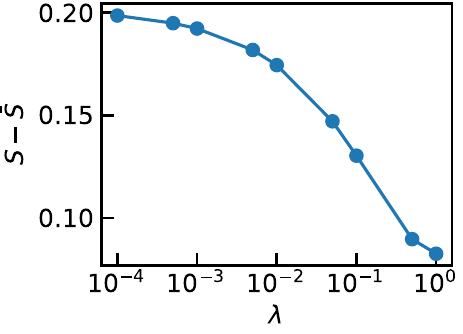}
    \caption{Zero-temperature reduction of the ordered moment as a function of anisotropy parameter $\lambda$. The value at $\lambda \to 0$ agrees with literature~\cite{manousakis1991rmp}.}
    \label{fig:moment_reduction}
\end{figure}

The longitudinal spin operator is: 
\begin{align}
    \widehat{S^z_{\mathbf{k}}} &= \frac{1}{\sqrt{N}}\sum_{j=1}^N e^{i({\mathbf k} + {\mathbf Q})\cdot {\mathbf R}_j} \widehat{S_j^z} \\
    &= \frac{1}{\sqrt{N}}\sum_{j=1}^{N/2} e^{i{\mathbf k}\cdot {\mathbf r}_j} (\widehat{S_{j,A}^z} - \widehat{S_{j,B}^z}) \\ 
    &= \sqrt{N} S \delta_{k,0} - \frac{1}{\sqrt{N}}\sum_{\mathbf{q}} \left(  a_{\mathbf{q}+\mathbf{k}}^\dagger a_{\mathbf{q}} + b_{-\mathbf{q}-\mathbf{k}}^\dagger b_{-\mathbf{q}} \right) \\ 
    &\approx \sqrt{N} S \delta_{k,0} - \frac{1}{\sqrt{N}}\sum_{\mathbf{q}} \Bigg( 
    (u_{\mathbf{q}+\mathbf{k}}u_{\mathbf{q}}+v_{\mathbf{q}+\mathbf{k}}v_{\mathbf{q}})(\alpha_{\mathbf{q}+\mathbf{k}}^\dagger \alpha_{\mathbf{q}} + \beta_{-\mathbf{q}-\mathbf{k}}^\dagger \beta_{-\mathbf{q}}) + 2\delta_{k,0}v_{\mathbf{q}+\mathbf{k}}v_{\mathbf{q}}  \nonumber \\ 
    & \qquad + u_{\mathbf{q+k}}v_{\mathbf{q}} ( \alpha^\dagger_{\mathbf{q+k}} \beta^\dagger_{-\mathbf{q}} + \beta_{-\mathbf{q}-\mathbf{k}}^\dagger \alpha_{-\mathbf{q}}^\dagger )
    + v_{\mathbf{q+k}}u_{\mathbf{q}} ( \alpha_{\mathbf{q+k}} \beta_{-\mathbf{q}} + \beta_{-\mathbf{q}-\mathbf{k}} \alpha_{\mathbf{q}} )
    \Bigg)
\end{align}
The zero-temperature reduction of the ordered moment is defined as
\begin{align}
S-\bar S &\equiv \int_{BZ}\frac{d^3q}{(2\pi)^3}\,v_{\mathbf q}^2 \\
&=
\int_{BZ} \frac{d^3 q}{(2\pi)^3}
\frac12\left(\frac{1}{\sqrt{1-\gamma_{\mathbf q}^2}} -1\right) \\
&\approx \int_{\Lambda} \frac{2\pi q_\parallel dq_\parallel dq_z}{(2\pi)^3} \frac{\sqrt{2+\lambda}}{\sqrt{q_\parallel^2+\lambda q_z^2}} 
+ \int_{BZ \backslash \Lambda} \frac{d^3 q}{(2\pi)^3} \frac12\left(\frac{1}{\sqrt{1-\gamma_{\mathbf q}^2}} -1\right) 
\end{align}
where the first term is an integral over a cylinder $k_\parallel, k_z \leq \Lambda$  and is finite: 
\begin{align}
    \int_{\Lambda} \frac{2\pi q_\parallel dq_\parallel dq_z}{(2\pi)^3} \frac{\sqrt{2+\lambda}}{\sqrt{q_\parallel^2+\lambda q_z^2}} 
    &= \frac{\Lambda^2\,\sqrt{2+\lambda}}{(2\pi)^2} \left[ \sqrt{1+\lambda}-\sqrt{\lambda}+\frac{1}{\sqrt{\lambda}}\operatorname{asinh}\!\big(\sqrt{\lambda}\big) \right]
    \sim \frac{\Lambda^2 \sqrt{2}}{2\pi^2}
\end{align}
Fig.~\ref{fig:moment_reduction} shows the zero-temperature reduction of the ordered moment as a function of anisotropy parameter $\lambda$.
At finite temperature, the thermal terms $(u^2_q+v^2_q)n_q \sim 1/q^2$ and $u_q v_q n_q \sim 1/q^2$ give finite corrections.

For $\mathbf k \neq 0$ the retarded spin susceptibility is:
\begin{align}
    \mathrm{Im} \chi^{zz}(\mathbf{k}, \omega) &= 
    \pi \frac{1}{N} \sum_{\mathbf{q}} \Big( (u_{\mathbf{q}+\mathbf{k}}u_{\mathbf{q}}+v_{\mathbf{q}+\mathbf{k}}v_{\mathbf{q}})^2 (n_B(\omega_{\mathbf{q}}) - n_B(\omega_{\mathbf{q}+\mathbf{k}}))  \delta(\omega + \omega_{\mathbf{q}} - \omega_{\mathbf{q}+\mathbf{k}}) \n \\ 
    & \quad + 4u_{\mathbf{q}+\mathbf{k}}u_{\mathbf{q}} v_{\mathbf{q}+\mathbf{k}}v_{\mathbf{q}} (1 + n_B(\omega_{\mathbf{q}}) + n_B(\omega_{\mathbf{q}+\mathbf{k}})) \delta(\omega - \omega_{\mathbf{q}} - \omega_{\mathbf{q}+\mathbf{k}}) \Big) 
\end{align}
Therefore, the normalized QFI density for alternating $S^{z}$ is:
\begin{align}
    \mathrm{nQFI}(\widehat{S^{z}},{\mathbf k},T) &= \frac{1}{4S^2} \frac{4}{\pi} \int_{0}^{\infty} d\omega \tanh\left( \frac{\beta \omega}{2} \right) \text{Im} \chi^{zz}(\mathbf{k}, \omega) \\
    &=\frac{1}{S^2}\frac{1}{N}\sum_{\mathbf q}\Big[
    |B_{\mathbf q,\mathbf k}|^2\,(1+n_{q+k}+n_q)\,
    \tanh\Big(\frac{\beta(\omega_{q+k}+\omega_q)}{2}\Big) \nonumber \\
    & \hspace{2cm} +|A_{\mathbf q,\mathbf k}|^2\,(n_q-n_{q+k})\,
    \tanh\Big(\frac{\beta(\omega_{q+k}-\omega_q)}{2}\Big)
    \Big].
\end{align}
where $A_{\mathbf q,\mathbf k}=u_{q+k}u_q+v_{q+k}v_q$, $B_{\mathbf q,\mathbf k}=u_{q+k}v_q+u_qv_{q+k}$, $n_q\equiv n_B(\omega_q)$.

At $\mathbf k=0$, the normalized QFI density in the longitudinal channel is therefore
\begin{align}
\mathrm{nQFI}(\widehat{S^{z}},\mathbf k=0, T)
&= \frac{1}{S^2} \frac{1}{N}\sum_{\mathbf q}
4 u_{\mathbf q}^2 v_{\mathbf q}^2 (1+2n_q) \tanh(\beta \omega_q) \\ 
&= \frac{1}{S^2} \Bigg( \int_{BZ\backslash\Lambda} \frac{1}{1/\gamma_{\mathbf q}^2 - 1} \frac{\tanh(\beta \omega_q)}{\tanh(\beta \omega_q/2)} + \int_\Lambda \frac{2+\lambda}{q_\parallel^2+\lambda q_z^2} 
\frac{\tanh(\beta \omega_q)}{\tanh(\beta \omega_q/2)} \Bigg) 
\end{align}
where $\frac{d^3q}{(2\pi)^3}$ is implicit in the integrals. 
Notice both regions around $(0,0,0)$ and $(\pi,\pi,\pi)$ are included in the cutoff cylinder $\Lambda$. 
 
In the $T\to \infty$ we have $\frac{\tanh(\beta \omega_q)}{\tanh(\beta \omega_q/2)} \approx 2$.
In the $T\to 0$ limit, we have $\frac{\tanh(\beta \omega_q)}{\tanh(\beta \omega_q/2)} \approx 1$.
Therefore, we have the relation between the longitudinal QFI at zero and infinite temperature:
\begin{align}
    \mathrm{nQFI}(\widehat{S^{z}},\mathbf k=0,T=0) = \frac12 \mathrm{nQFI}(\widehat{S^{z}},\mathbf k=0,T=\infty)
\end{align}

The integral over the cylinder $\Lambda$ is convergent 
\begin{align}
    I(T) &= \int_{-\Lambda}^\Lambda \frac{dq_z}{2\pi} \int_0^\Lambda \frac{2\pi q_\parallel dq_\parallel}{(2\pi)^2} \frac{2+\lambda}{q_\parallel^2+\lambda q_z^2} 
    \frac{\tanh(\beta \omega_q)}{\tanh(\beta \omega_q/2)} 
\end{align}
For $T=\infty$, the integral gives
\begin{align}
    I(T=\infty) &= \frac{(2+\lambda)\Lambda}{(2\pi)^2} \left[ \ln\left(\frac{1+\lambda}{\lambda}\right) + \frac{2}{\sqrt{\lambda}}\arctan(\sqrt{\lambda}) \right]
\end{align}

\section*{Spin wave analysis for $J_1$-$J_2$ model}\label{sec:spin_wave_J1J2}
\begin{figure}[th]
    \centering
    \includegraphics[width=0.5\linewidth]{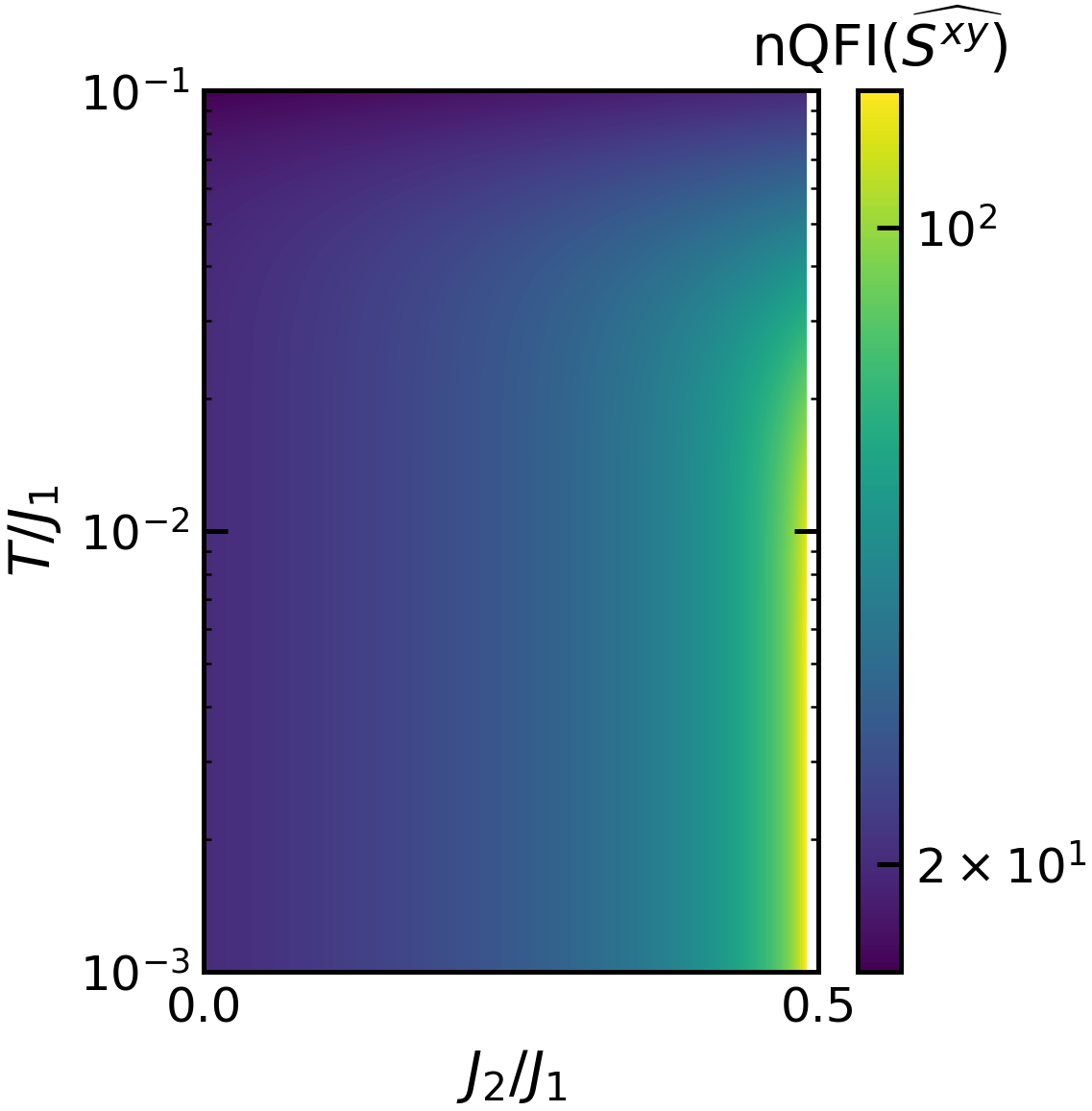}
    \caption{Normalized QFI density (nQFI) of $\widehat{S^x_i}$(or $\widehat{S^y_i}$) as a function of temperature $T$ and frustration parameter $J_2/J_1$. We set parameter $\sqrt{2}S|{\mathbf{q}}|=0.1$ at $\lambda=0$. }
    \label{fig:pd}
\end{figure}
Now we consider the spin-wave calculation for the $J_1-J_2$ model on a tetragonal lattice. 
In the N\'{e}el phase, the $J_2$ term is
\begin{align}
    H_2 &\approx z_2J_2 \bar{S} \sum_{\mathbf{k}}
    ( (\gamma_{2,\mathbf{k}} -1 ) a_{\mathbf{k}}^\dagger a_{\mathbf{k}} 
    + (\gamma_{2,-\mathbf{k}} -1 )  b_{-\mathbf{k}}^\dagger b_{-\mathbf{k}} )
\end{align}
where $\gamma_{2,\mathbf{k}}=\frac{1}{z_2}\sum_{\delta_2} e^{i {\mathbf k}\cdot\delta_2}=\frac{1}{2}[\cos(k_x+k_y)+\cos(k_x-k_y)]$ and the coordinate number is $z_2=4$.
The final linearized Hamiltonian takes the form
\begin{align}
    H = \sum_{\mathbf{k}} A_{\mathbf{k}}  a_{\mathbf{k}}^\dagger a_{\mathbf{k}} + A_{\mathbf{k}} b_{-\mathbf{k}}^\dagger b_{-\mathbf{k}} + B_{\mathbf{k}} a_{\mathbf{k}} b_{-\mathbf{k}} + B_{\mathbf{k}}^* a_{\mathbf{k}}^\dagger b_{-\mathbf{k}}^\dagger
\end{align}
where the coefficients are 
\begin{align}
    A_{\mathbf{k}} &= zJ_1\bar{S} + z_2J_2 \bar{S} (\gamma_{2,{\mathbf k}}-1) \approx zJ_1\bar{S} - z_2J_2 \bar{S} \frac{|\mathbf k|^2}{2}  \\ 
    B_{\mathbf{k}} &= zJ_1\bar{S} \gamma_{{\mathbf k}} \approx  zJ_1\bar{S} \left( 1- \frac{|\mathbf k|^2}{4}\right)
\end{align}
Therefore, the dispersion is 
\begin{align}
    \omega_{\mathbf{k}} &= \sqrt{A_{\mathbf{k}}^2 - B_{\mathbf{k}}^2} \\ 
    &= zJ_1\bar{S} \sqrt{ \left(1+\frac{z_2 J_2}{z J_1}(\gamma_{2,{\mathbf{k}}}-1) \right)^2-\gamma_{\mathbf{k}}^2 } \\ 
    &\approx \sqrt{2z} J_1 \bar{S} |\widetilde{\mathbf{k}}| \sqrt{1- \left[\frac{2z_2J_2}{zJ_1} - \left( \frac{z_2^2 J_2^2}{2 z^2J_1^2} - \frac18 \right)  k_\parallel^2\right]\frac{k_\parallel^2}{k_\parallel^2+\lambda k_z^2} } 
\end{align}
where $|\widetilde{\mathbf{k}}|=\sqrt{k_\parallel^2+\lambda k_z^2}$.
The spin wave gets flat at $J_1=2J_2$. 

We consider the case $\lambda \rightarrow0$.
The Bogoliubov transformation parameters are 
\begin{align}
    u_{\mathbf k} &= \sqrt{\frac{B_{\mathbf{k}}^2}{B_{\mathbf{k}}^2-(A_{\mathbf{k}}-\omega_{\mathbf{k}})^2}} 
    \\ 
    v_{\mathbf k} &= -\sqrt{\frac{(A_{\mathbf{k}}-\omega_{\mathbf{k}})^2}{B_{\mathbf{k}}^2-(A_{\mathbf{k}}-\omega_{\mathbf{k}})^2}} 
\end{align}
The prefactor in QFI is $(u_{\mathbf k}-v_{\mathbf k})^2 \approx \frac{2zJ_1 \bar{S}}{\omega_{\mathbf k}}$
and the final result is
\begin{align}
    \mathrm{nQFI}(\widehat{S^{xy}},{\mathbf k},T) 
    &\approx \frac{1}{4S^2} \frac{4\sqrt{2}\bar{S}}{|\mathbf{k}|} \tanh\left( \sqrt{2}\beta J \bar{S} |\mathbf{k}|\sqrt{1-\frac{2J_2}{J_1}} \right) \frac{1}{\sqrt{1-\frac{2J_2}{J_1}}}
\end{align}
At $\mathbf{k} = 0$, the normalized QFI density is:
\begin{equation}
    \mathrm{nQFI}(\widehat{S^{xy}},{\mathbf k} = 0,T) = 2 \beta J \left(\frac{\bar S}{S}\right)^2 
\end{equation}
At $T=0$, the normalized QFI density is:
\begin{equation}
    \mathrm{nQFI}(\widehat{S^{xy}},{\mathbf k},T=0) = \frac{\sqrt{2}\bar{S}}{S^2 |\mathbf{k}|} \frac{1}{\sqrt{1-\frac{2J_2}{J_1}}}
\end{equation}

Fig.~\ref{fig:pd} shows the temperature and frustration dependence of transverse nQFI at small but nonzero $\mathbf{q}$.
Transverse nQFI diverges at zero temperature for $\mathbf{q}=0$ or $J_2/J_1=0.5$.

\end{document}